\documentclass[iop]{emulateapj}
\usepackage{epsfig}
\usepackage{amsmath}
\usepackage{enumerate}

\usepackage{natbib}
 
\def\simlt{\lower.5ex\hbox{$\; \buildrel < \over \sim \;$}}
\def\simgt{\lower.5ex\hbox{$\; \buildrel > \over \sim \;$}}

\def\sT{\sigma_{\rm T}}

\def\beq{\begin{equation}}
\def\eeq{\end{equation}}
\def\ba{\begin{eqnarray}}
\def\ea{\end{eqnarray}}
\def\bB{{\,\mathbf B}}

\def\bj{{\,\mathbf j}}

\def\Rmax{R_{\rm max}}

\def\BQ{B_Q}

\def\M{{\cal M}}

\def\F{{\cal F}}

  \def\ang{\vartheta}
\def\angc{\tilde{\vartheta}}

\def\omegac{\tilde{\omega}}
\def\f{f_e}

\def\alf{\alpha}

\def\Sect{{\rm Section}}

\def\Rlc{R_{\rm lc}}
\def\mum{\hat{\mu}}

\def\Ethr{E_{\rm thr}}
\def\Ec{\tilde{E}}

\def\sigres{\sigma_{\rm res}}

\def\gsc{\gamma_{\rm sc}}

\def\const{{\rm const}}

\def\Lt{L}
\def\dNt{\dot{N}}

\def\ff{\digamma}

\def\Q{K}
\def\brad{b_0}
\def\Esc{E_{\rm sc}}
\def\gav{\bar{\gamma}}

\def\labs{l_{\rm abs}}

\def\Rs{R_\star}

\def\Etarget{E_t}
\def\ntarget{n_t}
\def\Lstar{L_{\rm th}}
\def\Drag{D}
\def\dNe{\dot{N}}
\def\Ninj{N_{\rm inj}}
\def\Emax{E_{\max}}

\def\phil{\phi_\star}
\def\thl{\theta_\star}

\def\Bpole{B_{\rm pole}}
\def\bk{\mathbf k}
\def\Ltail{L_{\rm tail}}
\def\Lann{L_{\rm ann}}
\def\rem{{\mathbf r}_{\rm em}}
\def\rb{{\mathbf r}_0}
\def\ymin{y_{\min}}
\def\ymax{y_{\max}}
\def\yres{y_{\rm res}}
\def\gampr{\gamma_p}
\def\gmin{\gamma_1}

\def\Eq{Equation}
\def\Eqs{Equations}

\newbox\grsign \setbox\grsign=\hbox{$>$} \newdimen\grdimen \grdimen=\ht\grsign
\newbox\simlessbox \newbox\simgreatbox \newbox\simpropbox
\setbox\simgreatbox=\hbox{\raise.5ex\hbox{$>$}\llap
     {\lower.5ex\hbox{$\sim$}}}\ht1=\grdimen\dp1=0pt
\setbox\simlessbox=\hbox{\raise.5ex\hbox{$<$}\llap
     {\lower.5ex\hbox{$\sim$}}}\ht2=\grdimen\dp2=0pt
\setbox\simpropbox=\hbox{\raise.5ex\hbox{$\propto$}\llap
     {\lower.5ex\hbox{$\sim$}}}\ht2=\grdimen\dp2=0pt
\def\simgt{\mathrel{\copy\simgreatbox}}
\def\simlt{\mathrel{\copy\simlessbox}}

\begin{document}

\title{On the mechanism of hard X-ray emission from magnetars}

\author{Andrei M. Beloborodov}
\affil{Physics Department and Columbia Astrophysics Laboratory,
Columbia University, 538  West 120th Street New York, NY 10027;
amb@phys.columbia.edu}

\begin{abstract}
Persistent activity of magnetars is associated with electric    
discharge that continually injects relativistic particles into the magnetosphere.
Large active magnetic loops around magnetars must be filled with outflowing 
particles that interact with radiation via resonant scattering and spawn 
electron-positron pairs. The outflow energy is processed into copious $e^\pm$ 
until the plasma enters outer parts of the loop where the magnetic field 
is reduced below $10^{13}$~G.
In the outer zone, photons scattered by the outflow do not convert to 
$e^\pm$ pairs and the outflow radiates its energy away. The escaping radiation
forms a distinct hard X-ray peak in the magnetar spectrum.
It has the following features:
(1) Its luminosity $L=10^{35}-10^{36}$~erg~s$^{-1}$ can easily exceed the 
thermal luminosity from the magnetar surface. 
(2) Its spectrum extends from 10~keV to the MeV band with a hard 
spectral slope, which
depends on the object inclination to the line of sight.
(3) The anisotropic  hard X-ray emission exhibits
 strong pulsations as the magnetar spins.
(4) The emission spectrum typically peaks around 1~MeV, but the peak position 
significantly oscillates with the spin period.
(5) The emission is dominated by the extraordinary polarization mode
at photon energies below $\sim 1$~MeV.
(6) The decelerated pairs accumulate and annihilate at the top of the 
magnetic loop, and emit the $511$-keV line with luminosity 
$\Lann\sim 0.1L$. Features (1)-(3) agree with available data;
(4)-(6) can be tested by future observations.
\end{abstract}



\section{Introduction}

Besides spectacular outbursts, magnetars produce persistent or 
decaying X-ray emission with luminosity 
$L\sim 10^{34}-10^{36}$~erg~s$^{-1}$.
Two peaks are observed in their X-ray spectra, with comparable luminosities.
The first peak is near 1~keV; it is associated with thermal emission from the 
neutron star surface. The second peak is above 100~keV. 
Its low-energy slope  (between 10 and 100~keV) was observed
in 7 magnetars\footnote{Non-detections of similar hard X-ray emission in 
      other magnetars are inconclusive because of insufficient sensitivity 
     (W.~Hermsen, private communication).} 
(Kuiper et al. 2008; Enoto et al. 2010), with a typical photon 
index $\Gamma\sim 1-1.5$. The emission shows pulsations with 
the rotation period; the pulsed fraction approaches 100\% at high energies. 

The present paper focuses on the mechanism of the hard X-ray component.
The puzzle of hard X-ray emission was previously discussed in several works 
(Thompson \& Beloborodov 2005; Heyl \& Hernquist 2005; 
Beloborodov \& Thompson 2007;  Baring \& Harding 2007; 
Lyubarsky \& Eichler 2008). 
In general, nonthermal luminosity from an isolated neutron star must be 
fed by some form of energy release in its magnetosphere. 
In magnetars, this process occurs in the closed magnetosphere via  
electric discharge (Beloborodov \& Thompson 2007). 
The created relativistic particles move along the magnetic field lines and 
may radiate their kinetic energy in two ways:

\begin{enumerate}

\item
Particles moving toward the neutron star will hit its surface and  
deposit their energy in a dense layer, 
which can radiate bremsstrahlung photons.

\item
Particles can radiate their energy in the magnetosphere by scattering 
photons around the star.

\end{enumerate} 

We consider in this paper magnetic loops that extend to radii $r\simgt 5R$, 
where $R$ is the radius of the neutron star (\Sect~2.1). 
We assume that relativistic particles are injected
in the low parts of the loops where the magnetic field is very strong, 
$B\simgt \BQ=m_e^2c^3/\hbar e =4.4\times 10^{13}$~G.
As explained below, the particles cannot immediately radiate their energy
via scattering. Instead, the energy is adiabatically processed into a standard
outflow along the magnetic field lines. This processing occurs via creation
of secondary $e^\pm$ pairs whose number grows by a factor of $\sim 10^2$
as the flow moves away from the star toward the top of the magnetic loop
(Figure~1). Pair loading ends where $B\approx 10^{13}$~G; at this 
point the outflow begins to lose energy to escaping photons.
We find that the outflowing plasma radiates all its energy in the zone 
$3\times 10^{11}<B<10^{13}$~G, in the upper parts of the loop.
The released radiation is practically independent of the details of energy 
injection near the star because of the energy processing in the pair-loading 
zone $B>10^{13}$~G.

\begin{figure}
\epsscale{1.0}
\plotone{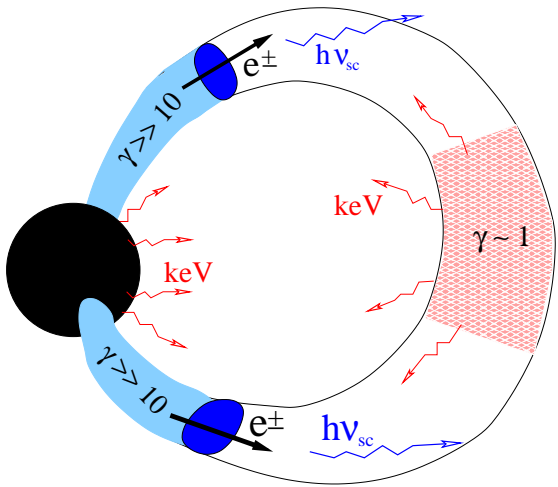}
\caption{
Sketch of an activated magnetic loop. Relativistic particles are 
injected near the star where $B>\BQ=4.4\times 10^{13}$~G. 
Large $e^\pm$ multiplicity $\M\sim 100$  (\Eq~\ref{eq:M}) develops in 
the adiabatic zone $B>10^{13}$~G (shaded in blue).
The outer part of the loop is in the radiative zone; here the scattered photons
of energy $E=h\nu_{\rm sc}$ escape and form the 
hard X-ray spectrum that is calculated in \Sect~3. The outflow decelerates
(and annihilates) at the top of the loop, shaded in pink;
here it becomes very opaque to the thermal keV photons flowing from the star.
Photons reflected from the pink region 
 have the best chance to be upscattered by 
the relativistic outflow in the lower parts of the loop, and control its
deceleration (\Sect~2.2).
}
\end{figure}

The paper is organized as follows. Section~2 describes the relativistic outflow,
its formation and dynamics. It is approximated by a simple analytical model, 
which is confirmed by detailed numerical 
simulations in the accompanying paper (Beloborodov 2012, hereafter B12).
Radiation emitted by the outflow is calculated in \Sect~3.
It is found to have an extended hard spectrum between 10~keV and $\sim 1$~MeV
and form a pronounced hard X-ray peak in the magnetar spectrum.
Section~4 explores the anisotropy and pulsations
of radiation produced by this mechanism.


\section{Relativistic flow in the j-bundle}

\subsection{Source of energy}

The radiative mechanism described in this paper is independent of
the energy source as long as the source generates relativistic 
particles near the magnetar where $B>10^{13}$~G.
We here briefly describe the expected source.

Like the sun, magnetars are believed to have 
twisted magnetospheres that are
deformed by surface motions  (Thompson et al. 2000; 2002).  
Beloborodov (2009, hereafter B09) developed electrodynamic theory   
for the dissipative, twisted magnetosphere attached to a conducting sphere
(neutron star). The theory predicts that  the magnetic twist concentrates on 
field lines that extend far from the conducting star, forming an extended 
bundle of electric currents. 
This ``j-bundle'' is heated by ohmic dissipation and emits radiation.  
The currents are nearly force-free, $\bj\times\bB=0$, and sustained by a 
longitudinal voltage $\Phi$ along the magnetic field lines. 
Net current circulating  through the magnetosphere, $I$, generates ohmic 
power $L=I\Phi$ that feeds the observed activity. 

Similar ohmic heating 
occurs in ordinary pulsars, and it may be useful to compare them with
magnetars. In ordinary pulsars, the magnetospheric twist is 
pumped at the light cylinder by the rotation of the star, 
and ohmic dissipation occurs on the {\it open} field lines. 
The electric current is then roughly given by
$I\sim c\mum/\Rlc^2$ (where $\Rlc$ is the light-cylinder radius
and $\mum$ is the magnetic dipole moment of the star),
and voltage $\Phi$ can exceed $10^{12}$~V. The dissipated power
$I\Phi$ is always smaller than the spindown luminosity of the star.
In magnetars, surface motions twist the {\it closed} magnetosphere.
Then the electric 
current may be as large as $I\sim c\mum/R^2$, where $R$ is the star radius.
The characteristic voltage is $\Phi\sim 10^9$~V 
(Beloborodov \& Thompson 2007).
The dissipated power $I\Phi$ is typically much larger than the spindown power. 

Ohmic dissipation tends to  gradually remove  electric 
currents from the closed magnetosphere as described in B09. 
This process creates a growing ``cavity'' with vanished current density $j=0$.
The currents have longest lifetimes on magnetic field lines 
with large apex radii $\Rmax\gg R$.
As a result, the magnetospheric activity becomes confined to the bundle 
of extended field lines. This theoretical picture agrees
with observations (see Beloborodov 2011 for a review). 

Consider magnetic field lines that extend sufficiently far from the star into 
the region where $B$ is smaller than a given value $B_1$. 
Luminosity that can be generated on
these field lines may be estimated as follows. Approximating magnetic 
field by a dipole configuration with a dipole moment $\mum$, one finds 
that the field lines reaching the region $B<B_1$ carry the magnetic flux
$\ff_1=2\pi \mum/R_1$ filling the hemisphere of radius 
$R_1=(\mum/B_1)^{1/3}$.
If the field lines are twisted with amplitude $\psi$, they must carry 
electric current (according to Stokes theorem, see B09 and Appendix~C),
\beq
\label{eq:I1}
   I_1\approx \frac{c\psi\ff_1}{8\pi R_1}=\frac{c\,\mum\,\psi}{4R_1^2}.
\eeq
The twist amplitude $\psi$ is measured in radians. It has the meaning of 
relative azimuthal position of the northern and southern footpoints of the 
magnetospheric field line. The generated power is $L=I_1\Phi$, which yields
\begin{eqnarray}
\nonumber
   L\approx 10^{36}\,\psi\left(\frac{\mum}{10^{33}{\rm ~G~cm}^2}\right)^{1/3}
                \left(\frac{\Phi}{4\times 10^9\rm~V}\right) \\
       \times
                \left(\frac{B_1}{10^{12}\rm ~G}\right)^{2/3} {\rm ~erg~s}^{-1}.
\end{eqnarray}
Strong twists tend to inflate the magnetosphere (Wolfson 1995; 
Parfrey et al. 2012a,b). As long as $\psi\simlt 1$, 
this effect is modest (it scales as $\psi^2$ at $\psi<1$, see B09), 
and the poloidal field may be well approximated by a dipole configuration 
with dipole moment $\mum$.

\subsection{Interaction of outflowing particles with radiation field}

Discharge with voltage $\Phi\sim 10^9$~V injects electrons (or positrons) 
with high Lorentz factors $\gamma\sim e\Phi/m_ec^2\sim 10^3$.
The relativistic plasma created near the star expands along the magnetic 
field and forms a relativistic outflow, resembling the outflow along open field 
lines in ordinary pulsars (however, here plasma moves along {\it closed} 
field lines and is trapped in the magnetic loops around the neutron star). 
Magnetars are hot and bright; their dense radiation 
exerts a strong drag on the magnetospheric particles and controls the 
outflow velocity. On the other hand, the plasma 
significantly changes the radiation field around the magnetar.
The interaction of the flowing $e^\pm$ plasma and radiation may be described 
as a self-consistent radiative transfer problem. This problem is solved numerically 
in the accompanying paper (Beloborodov, in preparation, hereafter B12). 
The results can be summarized as follows.

The outflow interacts with radiation via resonant scattering; other processes 
turn out unimportant. Consider an outflowing electron (or positron) 
with Lorentz factor $\gamma=(1-\beta^2)^{-1/2}$,  and a target photon of 
energy $\Etarget$ propagating at an angle $\ang$ with respect to the 
outflow direction. 
Resonant scattering can occur if the photon energy in the electron rest frame,
\beq
\label{eq:Ec}
    \Ec=\gamma(1-\beta\cos\ang)\Etarget,
\eeq
matches Landau energy $\hbar\omega_B$,
\beq
\label{eq:res}
    \Ec=\hbar\omega_B=b\,m_ec^2,  \qquad b\equiv\frac{B}{\BQ}.
\eeq
The relativistic outflow in the low parts of magnetic loops interacts
with thermal keV radiation that has been {\it reflected} by 
plasma trapped at the top of the magnetic loops (Figure~1). The outflow sees
the reflected photons as main targets because they propagate at most favorable 
angles for resonant scattering (Appendix~A).

Scattering of thermal keV radiation gives 
a huge energy boost to photons, on average by a factor $\sim \gamma^2$. 
The relativistic outflow experiences significant energy losses by scattering 
a tiny fraction of photons around the magnetar.
{\it The outflow adjusts so that its scattering rate 
is just enough for the self-consistent deceleration.}
In essence, the relativistic outflow moves fast enough to interact with 
photons of energy $E\sim 10kT$ (the low-density exponential tail of the 
thermal spectrum), and slow enough to not interact with the main peak of the 
thermal spectrum $E\sim 3kT$. This condition, together with 
\Eq~ (\ref{eq:res}), determines that the scattering plasma moves with
Lorentz factor 
\beq
\label{eq:gsc}
     \gsc\approx \frac{b\,m_ec^2}{10kT}.
\eeq

The self-regulation of Lorentz factor is a key feature of the outflow that allows 
one to calculate its emission spectrum (\Sect~3). 
This feature is illustrated in Figure~2 that shows 
the dynamics of a relativistic electron resonantly interacting with radiation. 
The electron was injected with Lorentz factor $\gamma=1000$ at 
$B=10\BQ$ and the figure shows its trajectory in the $B$-$\gamma$ plane.
As the electron moves outward along the magnetic field line, it scatters
photons and loses energy. Each scattering event is a step 
down in $\gamma$ at fixed B, and each free path between 
subsequent scatterings is a step down in $B$ at constant $\gamma$.
In this figure, the {\it mean} free path $\ell(b)$ is plotted between scattering
events; a full Monte-Carlo simulation will be shown in \Sect~3.

The average effect of scattering may be described as a drag force acting 
on the electron, $\F(B,\gamma)$. 
This force $\F$ is sensitive to the electron position in the $B$-$\gamma$ 
plane. As indicated in the figure, the drag is enormous above the 
electron trajectory and negligible below the trajectory. In essence,
the electron surfs the steep 
slope of $\F(B,\gamma)$ so that  it stays outside and around the strong-drag 
region of the $B$-$\gamma$ plane. 
The relevant dimensionless parameter measuring the drag effect
is roughly given by (see Appendix~A),
\beq
    \Drag=\frac{r\,\F}{\gamma\,m_ec^2}\sim  10\,\frac{r}{R}\,\frac{y^2}{e^{y}},
    \quad y=\frac{\hbar\omega_B}{\gamma(1-\beta\cos\ang_{\max})\,kT},
\eeq
where $\ang_{\max}$ is the maximum angle of target photons with respect 
to the electron velocity; $\cos\ang_{\max}=-0.5$ is assumed in Figure~2.
The electron surfs with $\Drag\sim |d\ln B/d\ln r|\sim 3$, which corresponds 
to $y=6- 7$ and $\gamma\approx 100\, b\,(kT/0.5{\rm ~keV})^{-1}$. 
This condition defines a line in the $B$-$\gamma$ plane that quite accurately 
describes the electron trajectory, confirming \Eq~(\ref{eq:gsc}).

\begin{figure}
\epsscale{1.2}
\plotone{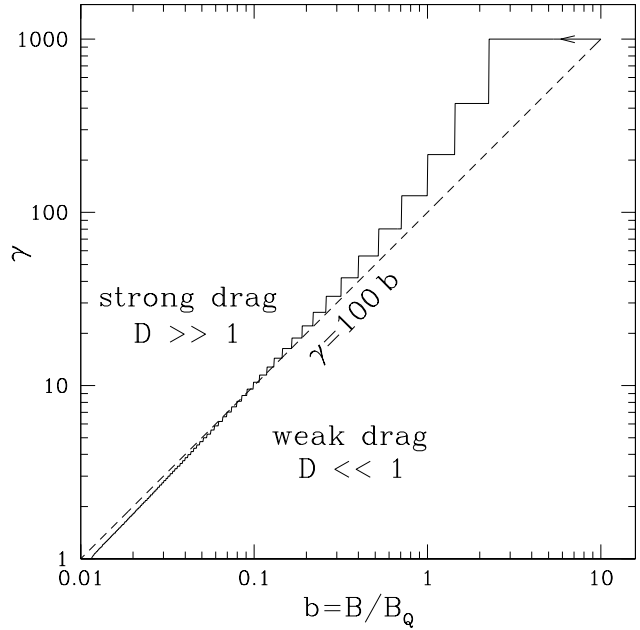}
\caption{
Deceleration of a relativistic electron moving along the magnetic field line
away from the star and interacting with ambient radiation field of temperature
$kT\approx 0.5$~keV (see text).
Each step down in $\gamma$ is a result of one scattering. 
}
\end{figure}

\bigskip

\subsection{Two zones of the outflow deceleration}

\subsubsection{Pair loading zone}

The relativistic outflow created in the ultra-strong magnetic field  must 
decelerate as it moves to weaker fields, according to \Eq~(\ref{eq:gsc}).
However, it is initially unable to radiate its energy away, because the 
scattered photons quickly convert to $e^\pm$ pairs that join the outflow.  
The deceleration occurs because the flow kinetic energy is shared by more
particles.

The mean expectation for the energy of a resonantly scattered photon is 
given by (see \Eq~\ref{eq:Eav})
\beq
\label{eq:Esc}
  E=\gsc\,b\,m_ec^2\,q(b),  \qquad 
  q=\frac{1}{b}\,\left(1-\frac{1}{\sqrt{1+2b}}\right),
\eeq 
which states that the photon energy 
in the electron frame, $\Ec=b\,m_ec^2$ (\Eq~\ref{eq:res}), is blueshifted in 
the lab frame by the typical Doppler factor $\sim\gsc$. The factor $q<1$ 
is a correction due to electron recoil in scattering.
In weak fields $q\approx 1$, and in strong fields ($b\gg 1$) 
$q\approx b^{-1}\ll 1$, consistent with the evident requirement $E<\gsc m_ec^2$. 
Combining \Eq~(\ref{eq:Esc}) with \Eq~(\ref{eq:gsc}), one finds that photons 
capable of converting to $e^\pm$ pairs, $E>2m_ec^2$, are generated in the 
region of rather strong magnetic field $b>b_0\sim 1/4$. 
A more accurate formula for $b_0$ is given in Appendix~B.  

Photons scattered in the zone $b>b_0$ do not escape. 
Their fate depends on the polarization state:
$\parallel$ photons directly convert to $e^\pm$, and 
$\perp$ photons split in two daughter $\parallel$ photons 
(see Appendix~A for the definition of $\parallel$ and $\perp$ polarization 
states). Conversion occurs 
immediately once the $\parallel$ photon satisfies the threshold condition,
\beq
\label{eq:thr}
   E>\Ethr=\frac{2m_ec^2}{\sin\ang}, 
\eeq
where $\ang$ is the photon angle with respect to the magnetic field.
The angle is initially small after scattering, $\ang\sim \gsc^{-1}$,
however it quickly grows as the photon propagates through the curved 
magnetic field. Therefore, almost all 
$\parallel$ photons of energy $E\gg 2m_ec^2$ produced in the zone 
$b>b_0$ are quickly absorbed. This fact is observed in our numerical 
simulations and discussed in Appendix~B
(cf. Medin \& Lai [2010], where a similar problem was studied for the 
polar cap near the magnetic axis).

In summary, the outflow radiation in the zone $b>b_0\sim 1/4$ is 
quickly absorbed and transformed into secondary $e^\pm$ pairs. This 
has a deceleration effect on the outflow, as the 
energy per particle is reduced (while the net energy of particles is 
conserved, since almost no photons escape this zone).

\subsubsection{Radiative zone}

As the outflow approaches $b_0$, the scattered photons begin to escape. 
In the region $b<0.1$ all scattered photons avoid conversion and splitting, 
and escape the magnetosphere, forming the hard X-ray component
of the magnetar spectrum. As the outflow reaches the region 
$b\sim 0.01$,  $\gsc$ decreases to a mildly relativistic value $\sim 1$ 
(see Equation~\ref{eq:gsc}), i.e. essentially all its energy has been 
radiated away.

We conclude that the relativistic outflow radiates almost all its energy in 
a well-defined region of magnetic field $0.01\simlt b\simlt 1/4$, 
regardless of the details of magnetospheric configuration --- the 
same conclusion is valid for a multipolar twisted magnetic field. 
\Sect~3 will describe a simple method to calculate the produced radiation.

\subsection{Multiplicity of $e^\pm$ pairs}

Let $2\dot{N}$ be the number flux of $e^\pm$ outflowing along 
both legs of the magnetic loop toward its top. Multiplicity is defined by
\beq
   \M=\frac{e\dot{N}}{I},
\eeq
where $I$ is the electric current circulating in the magnetic loop.
$\dot{N}$ grows in the pair-loading  zone $b>b_0$ to its final value that can 
be evaluated at $b\approx b_0$ using 
\beq
   2\dot{N}\,\gav(b_0)\, m_ec^2\approx L=\Phi I.
 \eeq
Here $\gav(b_0)$ is the average outflow Lorentz factor at $b\approx b_0$.
It is close to $\gsc$ given in \Eq~(\ref{eq:gsc}); $\gav\simlt\gsc$ is expected
if the particle momentum distribution has a significant width (see \Sect~2.5).
Then one finds,
\beq
\label{eq:M}
   \M\approx \frac{ e\Phi}{2\gav(b_0)\,m_ec^2} \sim 10^2,
\eeq
where we used the estimate $\gav(b_0)\simlt\gsc(b_0)\sim 20$.

The large multiplicity of created pairs implies that the outflow can easily 
conduct the electric current $I$ required by the twisted magnetic field
and screen electric fields. Therefore, no ``gaps'' with accelerating electric field 
are expected in the outer magnetosphere.

\subsection{Momentum distribution and collective effects}

A simplest model of $e^\pm$ outflow would assume that all particles have 
the Lorentz factor given by \Eq~(\ref{eq:gsc}).
In fact, the local distribution of Lorentz factor cannot be
a delta-function $\delta(\gamma-\gsc)$ ---
the real distribution must be broad. It is influenced by three effects:

\noindent
(1) Cold, single-fluid flow would be unable to carry the required electric current.
An electric field 
is induced in the magnetosphere to separate the average velocities of the 
positive and negative charges, so that the electric current is 
organized in the (nearly neutral) outflowing plasma (B12). It enforces
a minimum width of the momentum distribution of $e^\pm$. 

\noindent
(2) The cascade in the pair-loading zone injects particles with a broad 
range of momenta (see \Sect~3.2 below).

\noindent
(3) The momentum distribution of created particles $\f(p)$ is prone to 
two-stream instability. The instability tends to thermalize the distribution and 
reduce the gradients $d\f/dp$ that are responsible for the instability. 

Remarkably, it is possible to evaluate the outflow radiation 
even without knowing exactly how $\f(p)$ is
changed by collective electric fields. This is possible because
the outflow with a broad distribution function still must scatter 
radiation with Lorentz factor $\gsc$ given by \Eq~(\ref{eq:gsc}),
and the scattering rate is still dictated by how quickly the outflow must 
decelerate, keeping the particles at $\gamma\simlt\gsc$ and avoiding 
excessive drag. The surfing described in the end of \Sect~2.2 is still a 
valid description of the outflow dynamics.
A large width of $\f(p)$ only implies that at any given location
{\it a fraction} of particles actually scatter radiation --- those particles
that have $\gamma\approx\gsc$ --- while all other particles do not
have sufficiently high Lorentz factors for resonant scattering. 
The energies of scattered photons $\Esc$ and 
the net scattering rate are not much different from the single-fluid
model. This fact is observed in the simulations of B12, where $\f(p)$ is broad.


\section{Calculation of hard X-ray emission}

We build the emission model starting from the simple consideration of a single 
particle. Then we consider emission from the $e^\pm$ cascade that 
develops along the magnetic loop, without taking into account collective
effects. Finally, we formulate an approximate model for the $e^\pm$ outflow
with collective behavior, which will be used to calculate sample spectra in 
\Sect~4.

\subsection{Emission from one particle}

Consider the simplest case of one particle shown in Figure~2.
The particle is injected with a high Lorentz factor in the low part of the 
magnetic loop and decelerates as it moves away from the star.
What is the net spectrum of radiation emitted by the particle as it 
decelerates to $\gamma\sim 1$? 

The Lorentz factor $\gamma$ decreases along the 
``scattering curve,'' $\gamma\approx\gsc$,
which is given by \Eq~(\ref{eq:gsc}). 
It will be convenient to rewrite \Eq~(\ref{eq:gsc}) in the form,
\beq
\label{eq:gsc1}
    \gsc(b) = \Q\, b,    \qquad 
    \Q\approx 10^2 \left(\frac{kT}{0.5\rm ~keV}\right)^{-1}.
\eeq
The particle energy is passed to high-energy photons that are generated by 
scattering thermal radiation.
Each scattered photon has energy $\Ec=b\,m_ec^2$ in the particle rest frame,
as the scattering is {\it resonant}.\footnote{
     We consider the radiative zone where $b\ll 1$ and therefore neglect 
     corrections due to electron recoil.} 
Its blueshifted energy in the observer frame $E$ depends on the scattering 
angle. A simplest estimate for the emitted spectrum is obtained if we neglect 
the dispersion in scattering angles and take $E$ equal to its average value 
$\gamma\Ec$,
\beq
\label{eq:E1}
     E=\gamma\,b\,m_ec^2.
\eeq
Since the particle moves with $\gamma\approx\Q b$ (Figure~2),
we can substitute $b=\gamma/\Q$ in \Eq~(\ref{eq:E1}) and obtain the 
relation between the emitted photon energy $E$ and the particle energy 
$\gamma m_ec^2$,
\beq
\label{eq:rad}
   \gamma\, m_ec^2=\left(\Q \,m_ec^2 E\right)^{1/2}.
\eeq
Energy $dW_1=d\gamma\, m_ec^2$ is radiated in photons of energy $E$,
and the net emitted spectrum is given by
\beq
 \label{eq:dW1dE}
    \frac{dW_1}{dE}=\frac{d\gamma}{dE}\,m_ec^2
      =\frac{1}{2}\,\left(\frac{\Q\, m_ec^2}{E}\right)^{1/2}.
\eeq
Since $E\propto b^2$, the most energetic photons are emitted at the highest
$b\approx b_0$, near the boundary of the radiative zone, with energies
comparable to 1~MeV. As the particle moves to smaller magnetic fields $b$
and its Lorentz factor decreases, softer photons are produced,
forming the spectrum $\propto E^{-1/2}$.

\subsection{Emission from $e^\pm$ cascade (no collective effects)}

Again consider one high-energy particle injected with a Lorentz factor 
$\gampr>100$ at $b>1$. Emission from the particle in the 
radiative zone $b<b_0$ was evaluated in \Sect~3.1. 
Before emitting the escaping radiation, the particle scatters photons
in the zone $b>b_0$ and thus generates secondary $e^\pm$ pairs.
The secondary particles  also scatter photons, which can convert to more 
pairs. All $e^\pm$ pairs generated in this cascade enter the radiative zone
and eventually decelerate to $\gamma\sim 1$ as they outflow along the 
magnetic loop. What emission should be observed from the cascade?

In this section, we adopt the assumption that all particles flow freely  
and change their $\gamma$ only when they scatter a photon. This model 
ignores the interaction between particles via collective electric fields (\Sect~2.5). 
The opposite case of strong collective effects will be considered in \Sect~3.3.

\begin{figure}
\epsscale{1.2}
\plotone{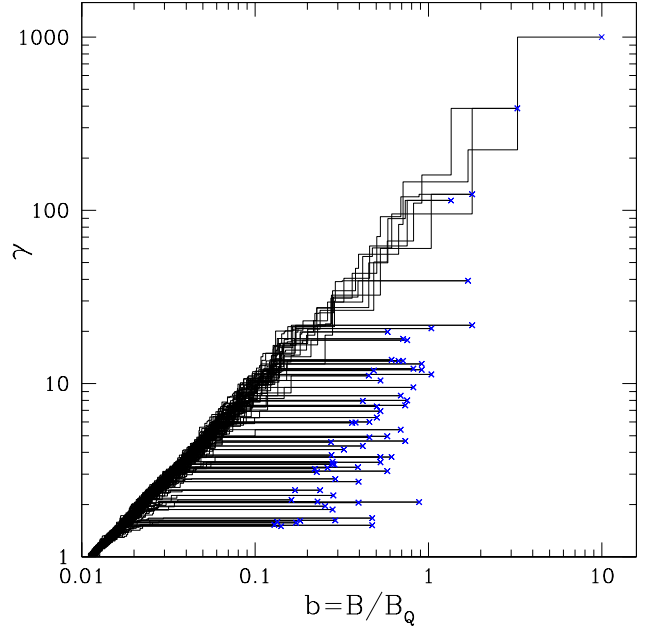}
\caption{
One random realization of $e^\pm$ cascade produced by one electron
injected with Lorentz factor $\gampr=10^3$ at magnetic field $B=10\BQ$.
136 secondary particles are created in this cascade; the injection points 
for the primary particle and 68 secondary pairs are shown by blue crosses. 
After injection, each particle moves toward and then along the 
``scattering curve,'' which is approximately described by \Eq~(\ref{eq:gsc}).
}
\end{figure}

The development of the cascade is shown in Figure~3; the details of
the calculation are given in Appendices~A and B.
Each created particle follows a track in the $B$-$\gamma$ 
plane similar to that in Figure~2, except that the secondary particles 
are injected with smaller energies. They start their tracks significantly below 
the scattering curve $\gsc=\Q b$ and have a long free path with 
$\gamma=\const$ before reaching the curve; then the particles begin to 
frequently scatter radiation and surf along the curve as shown in \Sect~2.2.

The cascade develops in the zone of a strong magnetic field 
where the photon splitting rate is high (Appendix~B).
The scattered photon can be in one of the two polarization states, $\perp$ 
or $\parallel$. The energetic $\perp$ photon
quickly splits in two daughter photons with $\parallel$ polarization, before 
  it gets a chance to convert to $e^\pm$. 
The $\parallel$ photons do not split;  they convert to $e^\pm$. 
The conversion occurs immediately after the propagating photon 
meets the threshold condition (\ref{eq:thr}). Therefore, $e^-$ and $e^+$ are 
created with total energy $E_\pm\approx \Ethr$, in the ground Landau state, 
and do not produce synchrotron emission.

Note that pair creation takes place in the region $b>0.1$ where 
photons should convert to positronium rather than free $e^\pm$ 
(Shabad \& Usov 1986).
The binding energy of positronium atoms is $\tilde{E}_b\sim 0.1$~keV,
and they are easily unbound by photons of energy comparable to 
$\tilde{E}_b$ (in the atom rest frame), i.e. by UV photons of energy 
$\sim 0.1\gamma^{-1}$~keV in the lab frame,
where $\gamma=E_\pm/2m_ec^2$ is the Lorentz factor of the 
positronium atom.
The rate of this reaction in a thermal radiation field was evaluated by 
Bhatia et al. (1988). In our problem, 
the positronium Lorentz factors are modest, and the resulting ionization 
rate is so high that essentially all $e^\pm$ pairs must be unbound.
Even if a positronium atom reaches the scattering curve before 
ionization, $e^+$ and $e^-$ begin to resonantly interact with X-rays as if
they were free particles (because $\hbar\omega_B\gg\tilde{E}_b$) and 
immediately become unbound due to recoil in scattering.

\begin{figure}
\epsscale{1.15}
\plotone{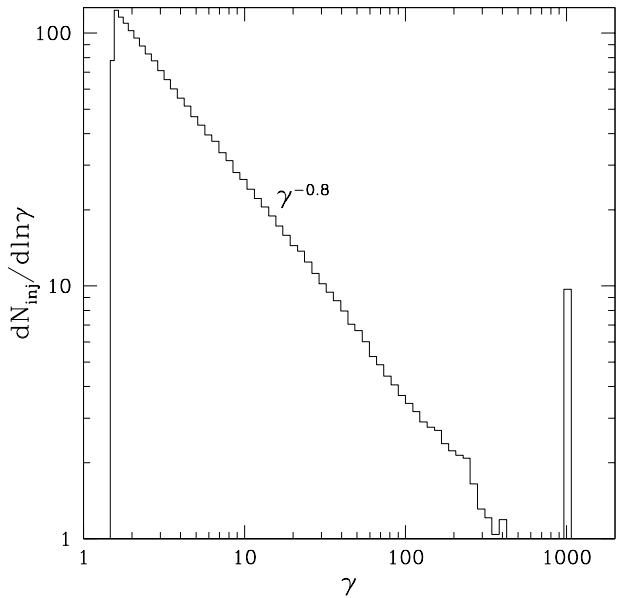}
\caption{Injection Lorentz factor distribution of $e^\pm$ pairs in 
cascades generated by one particle with $\gampr=10^3$
(cf. blue crosses in Figure~3, which shows one realization of the cascade).
The injection of the primary particle is shown by the peak at 
$\gamma=10^3$ (normalized to unity), and the broad smooth distribution
shows the injection of secondary $e^\pm$. Straight line shows the 
approximation given in \Eq~(\ref{eq:inj}).
}
\end{figure}

Our cascade simulation shows that the total number of secondary $e^\pm$ 
pairs (generated per one electron injected with Lorentz factor 
$\gampr=10^3$)  is $\Ninj\approx10^2$, similar to estimate (\ref{eq:M}).
The particle Lorentz factors at injection points (blue crosses in Figure~3)
form a distribution $d\Ninj/d\gamma$ that can be accurately calculated by 
simulating a large number ${\cal N}$ of random realizations of the cascade.
We use ${\cal N}=10^4$ to accumulate sufficient statistics. The obtained
distribution $d\Ninj/d\ln \gamma$ is shown in Figure~4. It is very well 
approximated by a simple power law,
\beq
\label{eq:inj}
  \frac{d\Ninj}{d\ln \gamma}=A\,\gamma^{-q}, \qquad \gamma>\gmin.
\eeq 
Here $\gmin=E/2m_ec^2\sim 1-2$ corresponds to photons of minimum energy 
$E$ that are able to convert to $e^\pm$; in our sample model $\gmin=1.5$.
The slope of the distribution in Figure~4 is $q\approx 0.84$, and the 
corresponding normalization factor $A$ is given by 
$A=q\gmin^{q}\Ninj\approx 0.86\Ninj$.

A cascade is characterized by its ``pair yield'' $Y$. It is defined as the mean 
expectation for the ratio of the total rest-mass energy of created pairs, 
$\Ninj m_ec^2$, to the energy of the primary particle $\gampr m_ec^2$. 
We find
\beq
\label{eq:Y}
  Y=\frac{\Ninj}{\gampr}=0.15.
\eeq

\begin{figure}
\epsscale{1.15}
\plotone{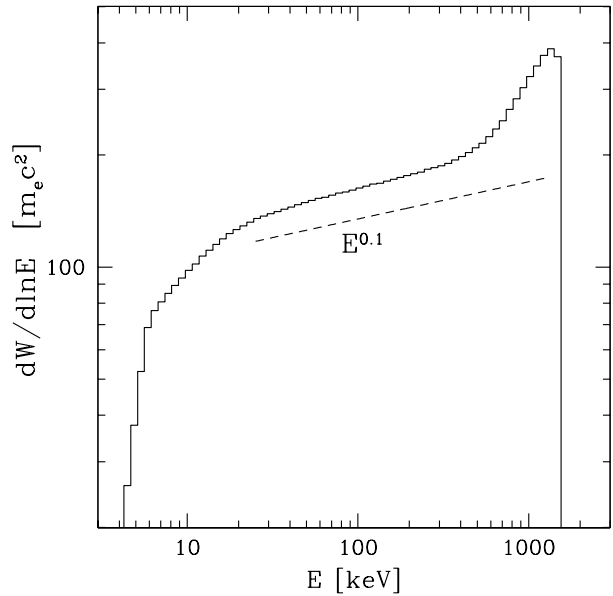}
\caption{Average spectrum of escaping radiation from cascades
generated by primary particles with Lorentz factor $\gampr=10^3$.
The total emitted energy $W$ equals the injected energy 
$\gampr m_ec^2$. Dashed line shows the analytical estimate
(\Eq~\ref{eq:dWdE}). Photon splitting creates an additional bump that 
is visible at $E\sim 1$~MeV. 
The bump consists of photons with $\parallel$ polarization. 
Emission at lower energies $E\ll 1$~MeV is dominated by photons with 
$\perp$ polarization.
}
\end{figure}

The same Monte-Carlo simulation gives the spectrum of radiation 
emitted by the cascade. We accumulate the statistics of all escaping 
photons (which avoid splitting and conversion) in ${\cal N}=10^4$ 
realizations of the cascade, and find their average spectrum. The result 
is shown in Figure~5. 
In this simulation, we assume that
radiation is produced by a thin magnetic tube for which 
the highest energy of escaping photons $E_{\max}=3m_ec^2$ 
(these photons are emitted at $b\approx b_0\approx 0.3$); 
therefore the obtained spectrum has a sharp break at $E\approx 1.5$~MeV.
We also performed more detailed simulations that track all photons as they 
propagate through the magnetosphere and check for absorption or 
splitting along their trajectories (Appendix~B); this gave a similar break at 
$E_{\rm max}$ whose exact value (comparable to $3m_ec^2$) depends 
on the radius of curvature of the magnetic tube.
  
A simple analytical estimate for the spectral slope below 1~MeV can 
be derived as follows. Each injected particle begins to radiate when 
it reaches the scattering curve $\gamma=\Q b$. Each particle surfing
along the curve emits the same standard spectrum $dW_1/dE$ given by 
\Eq~(\ref{eq:dW1dE}). Hence radiation produced at a given photon
energy $E$ is given by
\beq
\label{eq:WE}
    \frac{dW}{dE}=\frac{dW_1}{dE}\,
       \int_{\gamma_E}^{\gampr}\frac{d\Ninj}{d\gamma}\,d\gamma, 
  \qquad \gamma_E=\left(\frac{\Q\, E}{m_ec^2}\right)^{1/2},
\eeq
where $\gamma_E$ is the Lorentz factor of particles 
that emit at energy $E$ (\Eq~\ref{eq:E1}).
From \Eqs~(\ref{eq:WE}) and (\ref{eq:inj}) we find
\beq
\label{eq:dWdE1}
   \frac{dW}{d\ln E}= \frac{A}{2q}\,m_ec^2
                      \left(\frac{\Q\,E}{m_ec^2}\right)^{\frac{1-q}{2}}, 
   \qquad \frac{E}{m_ec^2}> \frac{\gmin^2}{\Q},
\eeq
which gives
\beq
\label{eq:dWdE}
  \frac{dW}{d\ln E} \approx 0.1 \gampr m_ec^2\,
              \left(\frac{\Q\,E}{m_ec^2}\right)^{0.1},
\eeq
where we used \Eq~(\ref{eq:Y}), $A/q=\Ninj\gmin^q$, and $q\approx 0.8$.
This expression only describes radiation produced by scattering; it
does not take into account photon splitting. It gives a reasonable 
approximation to the emitted spectrum at $E<m_ec^2$. 
The splitting of high-energy 
photons with the $\perp$ polarization, $\perp\rightarrow \parallel + \parallel$,
creates additional photons that form a bump near $E\sim 1$~MeV (Figure~5). 

\begin{figure}
\epsscale{1.2}
\plotone{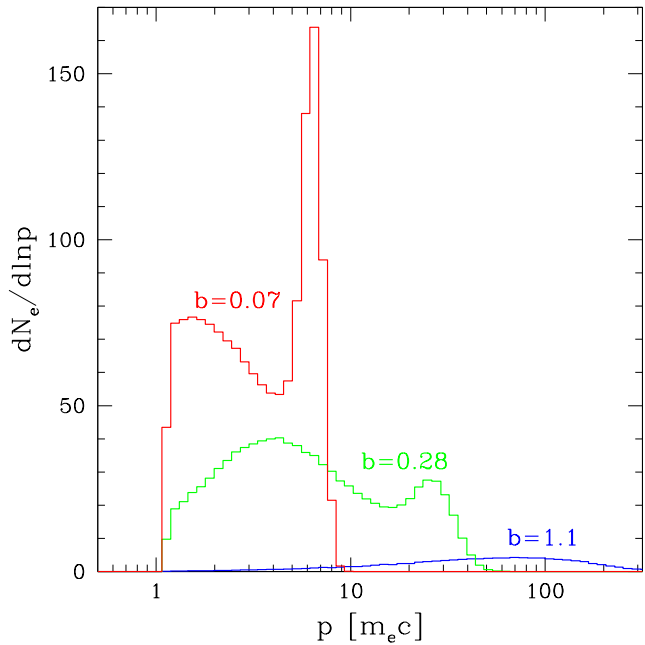}
\caption{Momentum distribution of outflowing $e^\pm$ pairs created by the 
cascade, without taking into account collective effects. The distribution 
changes along the magnetic tube with decreasing $b=B/\BQ$ as more 
particles are created and the particles lose momentum to scattering.
}
\end{figure}

The cascade model described above has one feature that makes the 
model questionable: the momentum distribution of $e^\pm$
develops two peaks and becomes unstable. 
We have calculated the distribution of $e^\pm$ created by 
the cascade and how it evolves with $b$ as the flow moves 
along the tube. The momentum distribution at three locations in the tube is 
shown in Figure~6. One can see the development of two peaks.
The low-energy peak ($p\approx 1.5$ at $b=0.07$) is formed by recently 
injected secondary particles. The high-energy peak ($p\approx 7$ at $b=0.07$)
is formed by older particles that were injected at higher $b$ and are now 
moving along the scattering curve (cf. Figure~3); its position evolves with $b$ 
according to $p\approx\gsc\approx 100 b$ (\Eq~\ref{eq:gsc1}).

The two-peak distribution is prone to plasma instability that excites 
Langmuir waves in the outflow.
We conclude that the model without collective effects is unstable, and 
the excitation of collective electric fields is inevitable,
which will change the distribution function. 
Below we will seek a simplest emission model for the $e^\pm$ 
outflow with collective behavior.

\subsection{Emission from $e^\pm$ outflow with collective behavior}

At a given location in the active loop, only a small fraction of particles 
have $\gamma\approx\gsc$ that is high enough to resonantly scatter radiation; 
most particles have $\gamma<\gsc$ and do not interact with radiation. 
This fact is true in the absence of collective effects (see Figure~3) and 
remains true in collective models (cf. numerical simulations in B12).
Since radiation is scattered by particles with known Lorentz factor 
$\gamma\approx \gsc$ (\Eq~\ref{eq:gsc1}), the details of $\f(p)$ are not 
important for the produced emission. It turns out that it is sufficient to know the 
{\it average} Lorentz factor $\gav(b)$,
\beq
   \gav=\int (1+p^2)^{1/2} \f(p)\,dp.
\eeq
Our model of $e^\pm$ outflow with collective behavior will avoid the 
complicated calculation of $\f(p)$; instead, we focus on $\gav$ and 
use the following approximation,
\beq
\label{eq:appr}
   \gav\propto \gsc,
\eeq
where $\gsc$ is given by Equation~(\ref{eq:gsc1}). 
In essence, $\gav(b)$ is kept somewhat below $\gsc(b)$ --- the function 
$\gsc(b)$ shows how the outflow must decelerate as it enters 
weaker magnetic fields. 
As an example of collective behavior, one can picture an outflow whose
distribution function maintains a constant shape, which shifts to lower 
$\gamma$ as the outflow moves to weaker magnetic fields.
Approximation~(\ref{eq:appr}) greatly simplifies the calculation of generated 
radiation. Remarkably, the details of the cascade, its multiplicity, and the coefficient of 
proportionality in \Eq~(\ref{eq:appr}) turn out unimportant --- they drop out 
from the calculation as will be seen below.
  
Consider a thin active magnetic flux tube.
We wish to know how its luminosity $L$ is distributed over 
$b$ along the tube, as this will determine the spectrum of produced radiation. 
We will use the simple fact that the power emitted by the tube element of 
length $dl$ equals the power lost by the outflow, i.e. the emitted power is 
determined by the deceleration law.
  
In the radiative zone, where no new $e^\pm$ pairs are created, the particle
number flux $\dNt$ is constant along the tube up to its summit, where particles
annihilate. The outflow power $2\dNt \gav m_ec^2$ 
(factor of 2 for two legs of the tube) decreases proportionally to $\gav$. 
The lost power transforms to radiation, and hence the produced luminosity 
is distributed over $b$ according to 
\beq
\label{eq:Lb}
   \frac{d\Lt}{db}=2\dNt \frac{d\gav}{db}\,m_ec^2=\const.
\eeq
Here $b=B/\BQ$ is used as a coordinate along the magnetic flux tube;
expression on the left side is constant along the tube 
because $\gav$ is proportional to $\gsc$ (\Eq~\ref{eq:appr}) and $\gsc$ 
is proportional to $b$ (\Eq~\ref{eq:gsc1}), so $d\gav/db=\const$.
The total power $L$ (emitted in the radiative zone $b<b_0$) equals
$(d\Lt/db)b_0$, which gives
\beq
\label{eq:dLdb}
   \frac{d\Lt}{db}=\frac{\Lt}{\brad}.
\eeq

Equations~(\ref{eq:gsc1}) and (\ref{eq:dLdb}) lead to a complete description
of the emitted radiation, as at every $b$ they determine the Lorentz factor of 
the emitting particles $\gsc$ and the emission rate.
Consider first the net spectrum of radiation emitted by the tube in all directions.
Similarly to \Sect~2.1,  a simplest estimate is obtained if we neglect the 
dispersion in scattering angles and take photon energy $E$ equal to its 
mean expectation $\gsc b\,m_ec^2$,
\beq
\label{eq:E}
     E=\Q\,m_ec^2\,b^2.
\eeq
Then one finds $db=(1/2)(\Q\, m_ec^2\, E)^{-1/2}\,dE$, and substitution to 
\Eq~(\ref{eq:dLdb}) gives the spectrum emitted by the tube,
\beq
\label{eq:LE}
   \frac{d\Lt}{dE}=\frac{\Lt}{2\,b_0\,(\Q\, m_ec^2 E)^{1/2}}.
\eeq
The total produced luminosity is proportional to $\int E^{-1/2}\,dE$ and peaks at 
the high-energy end as $E^{1/2}$.  The most energetic escaping photons are 
emitted at $b\approx b_0$, near the boundary of the radiative zone, with 
energies comparable to 1~MeV.  \Eq~(\ref{eq:LE}) does not include the
additional component that is generated near 1~MeV by photon splitting;
it will be included in the detailed Monte-Carlo models presented below.

In \Sect~2.3, the cascade without collective effects gave a softer spectrum
$dL/dE\propto E^{-0.9}$ (\Eq~\ref{eq:dWdE}). In that case, $\gav(b)$ was not 
proportional to $\gsc$; the secondary particles 
with $\gamma<\gsc$ kept constant $\gamma$ until 
they reached the scattering curve at lower $b$ and radiated their energy 
in softer photons. By contrast,
the collective behavior with $\gav\propto\gsc$ invokes a quick redistribution 
of energy between the injected particles with $\gamma<\gsc$ and the radiating 
particles with $\gamma\approx \gsc$.

\subsection{Detailed calculation: angular distribution}

To evaluate the emission received from an active magnetosphere by 
observers with various lines of sight, we need a more detailed calculation. 
A concrete geometry of the j-bundle must be chosen in such simulations.

One can view the active part of the magnetosphere (j-bundle) as a 
collection of infinitesimally thin active magnetic flux tubes. 
The flux tubes are assumed to extend sufficiently far from the 
neutron star, reaching the region of $B<10^{12}$~G. Consider one 
elementary tube that contains magnetic flux $\delta\ff$ and electric current 
$\delta I$. The discharge operates in the tube under voltage $\Phi$ and 
creates an outflow with power,
\beq
\label{eq:Lt}
    \delta L=\chi\,\Phi\,\delta I.
\eeq
Here we allowed a fraction $1-\chi$ of the total ohmic power released in 
the discharge zone to go to the footpoints of the tube, and the fraction 
$\chi$ is given to the outflowing particles. 

Emission from each piece of the tube is beamed along the 
outflow, and the emerging radiation is strongly anisotropic. 
Consider a short piece of the magnetic tube of length $dl$ which 
corresponds to $db=(db/dl)\,dl$. Its contribution to the tube luminosity is 
$(\delta L/2b_0)\,|db|$ (see \Eq~[\ref{eq:dLdb}] and take into account 
that there are two elements $dl$ that correspond to $db$, as the tube
has two footpoints).
The average energy of emitted photons is $\gsc\,b\,m_ec^2$, and hence
the photon emission rate for the tube element $dl$ is given by
\beq
   d(\delta\dot{N}_{\rm ph})=\frac{\delta L}{2\gsc b\,m_ec^2}\,\frac{|db|}{b_0}.
\eeq
It remains to determine the spectrum and polarization of the emitted photons.
In the radiative zone $b$ is small ($b\ll 1$) and hence electron
recoil per scattering is small. In this regime, the description of polarization and 
angular distribution of scattered photons is simple (see Appendix~A in B12 for 
details). Photon energy in the observer frame, $E$, is 
related to its emission angle in the electron rest frame, $\angc$, by
the Doppler transformation of $\Ec=\hbar\omega_B$,
\beq
\label{eq:Elab}
  E=\hbar\omega_B\,\gsc(1+\beta_{\rm sc}\cos\angc).
\eeq
 Approximately $3/4$ of scattered photons have
the $\perp$ polarization (cf.~\Eq~\ref{eq:polariz}); these photons 
are emitted with an isotropic distribution,
\beq
   P_\perp(\cos\angc)=\const=\frac{1}{2}.
\eeq 
The remaining 1/4 of photons are emitted with the $\parallel$ polarization, 
and their angular distribution is 
\beq
  P_\parallel(\cos\angc)=\frac{3}{2}\,\cos^2\angc.
\eeq
Using the Monte-Carlo technique, we generate 
the angular distributions for photons with $\perp$ and $\parallel$ 
polarizations, and determine the photon energies according 
to \Eq~(\ref{eq:Elab}). 

This straightforward Monte-Carlo simulation
gives the spectrum and angular distribution of emission from all parts
of the magnetic tube. We determine the fate of emitted 
photons by tracking their trajectories through the magnetosphere
and checking for splitting or conversion to $e^\pm$, which allows us to smoothly 
describe the transition from pair-loading to radiative zone (Appendix~B). 
Photons that avoid splitting and conversion form the observed spectrum. 
Finally, the net observed emission is obtained by integrating the emission along 
the tube and then taking the sum of emissions from all tubes in the j-bundle. 

The same calculation may be performed by integrating emission 
over the volume of the j-bundle in usual spherical coordinates.
This method is described in Appendix~C, where emissivity per unit volume 
$dL/dV$ is derived. Appendix~C also describes the concrete case of 
a twisted dipole magnetosphere.

\subsection{Multiple scattering of hard X-rays}

The Monte-Carlo simulation described above neglects the fact that 
the generated high-energy photons can scatter multiple times in the 
outflow before escaping. It turns out that this additional scattering weakly 
affects the emerging spectrum. Note two facts:
(1) Hard X-rays escape the scattering region on the light-crossing time
regardless of the number of scatterings; this is because the
plasma moves relativistically and 
the scattered photons are beamed along the flow, so they keep moving 
 away from the star.
(2) The outflow significantly loses energy only when it scatters {\it soft}
target photons that 
propagate at {\it large} angles $\ang$ with respect to the outflow direction;
then scattering boosts the photon energy by the large factor of 
$\sim \gsc^2\gg 1$.\footnote{
    This boost happens in the lab frame;  the photon energy is unchanged
    in the outflow rest frame --- only its angle is changed by scattering
    (from $\cos\angc\approx -1$ to a random value).}
Since the scattering reduces $\ang$ to $\sim\gsc^{-1}$
(which corresponds to a quasi-isotropic distribution of photons
in the outflow rest frame),
subsequent scatterings have a small effect on the photon energy.

To check the effect of multiple scattering on the hard X-ray spectrum
we used the Monte-Carlo transfer code developed in the accompanying 
paper (B12). We followed the propagation and scattering of the generated 
hard X-rays through the flowing magnetospheric plasma and found that 
multiple scattering changes the emerging spectrum by less than 20\%.
This change makes the spectrum slightly harder; 
it will be neglected in this paper.


\section{Sample numerical models}

\subsection{Model setup}

Consider a neutron star of radius $R=10$~km, temperature $kT=0.5$~keV,
and magnetic dipole moment $\mum=5\times 10^{32}$~G (which 
corresponds to the magnetic field at the pole $\Bpole=10^{15}$~G).
Poloidal magnetic field is well approximated as dipole
if the twist amplitude is moderate, $\psi\simlt 1$ (B09; Parfrey et al. 2012b). 
We will use spherical coordinates $r,\theta,\phi$ with the polar axis aligned 
with the dipole moment; $\theta=0$ on the axis.
  
Consider an activated magnetic loop (j-bundle) whose northern
footpoint on the star is defined by
\beq
\label{eq:foot}
   \frac{\thl}{2}<\theta<\thl, \qquad 0<\phi<\phil.
\eeq
In our sample models, $\sin^2\thl=0.1$; 
$\phil$ is left as a free parameter
($\phil=2\pi$ describes an axisymmetric j-bundle).
Practically the same results are obtained if the j-bundle 
includes the entire polar cap $0<\theta<\thl$, as the field lines with 
footpoints at $\theta<\thl/2$ contribute only a small fraction $\sim 2^{-4}$ 
to the emitted luminosity.\footnote{
      Electric current in the j-bundle $\theta<\thl$ with a given twist 
      $\psi$ scales as $I\propto \psi\sin^4\thl$ (Appendix~C).}
The active loop extends to radii $r\simgt R/\sin^2\thl\approx 10R$.
We wish to know the spectrum 
of radiation emitted by the loop, as viewed along any given line of sight.
    
The total luminosity of the loop is controlled by its twist  amplitude $\psi$ 
and the discharge voltage $\Phi$ (\Sect~2.1). It is approximately given by 
(see B09 and Appendix~C)
\beq
   L \approx \frac{c\,\Phi\,\mum\,\psi\,\sin^4\thl}{4\,R^2}\,
       \left(\frac{\phil}{2\pi}\right).
\eeq
The discharge occurs in both hemispheres, forming two outflows
that meet and annihilate at the equatorial plane of the magnetic dipole. 
Therefore, the total emitted power corresponds to $\Phi$ that is two times 
larger than the threshold for the discharge; $\Phi=10^9-10^{10}$~V is 
expected (Beloborodov \& Thompson 2007) and consistent with 
observations (Beloborodov 2011). 
It is convenient to define the parameter
\beq
    H\equiv \psi\,\left(\frac{\Phi}{3\times 10^9 \rm ~V}\right)
                  \,\left(\frac{\phil}{2\pi}\right) \left(\frac{R}{10\rm ~km}\right).
\eeq
Then the total power emitted by the loop may be written as
\beq
\label{eq:Ltot}
    L\approx 4\times 10^{35}\,H\,\left(\frac{\Bpole}{10^{15}\rm ~G}\right)
                     \left(\frac{\sin^4\thl}{10^{-2}}\right)
                    {\rm ~erg~s}^{-1}.
\eeq
The observed emission from the loop is calculated using the Monte-Carlo 
technique as described in \Sect~3.4. 
We assume that the $e^\pm$ outflow behaves collectively (\Sect~3.3),
as this model is both simpler and more 
realistic than the cascade without collective effects.

The emission from the loop is strongly anisotropic and hence emission
observed along a given line of sight  
will be modulated by the rotation of the object.
Let us choose the fixed (non-rotating) lab frame so that its $z$-axis is 
parallel to the spin axis, and the $xz$ plane contains the observer's line of 
sight. The instantaneous orientation of the magnetosphere with respect to 
the lab frame is described by three Euler angles $\alpha$, $\varphi$, $\zeta$.
Here $\alpha$ is the angle between the magnetic and rotation axes,
$\varphi$ is the angle of rotation about the spin axis, and
$-\zeta=\phi_{\rm nodes}$ is the azimuthal angle 
(measured in magnetic coordinates $r,\theta,\phi$)
of the line of nodes $\vec{\Omega}\times\vec{\mum}$.

Let $\beta$ be the angle between the line of sight and the rotation axis. 
The instantaneous observed emission is determined by 
four angles $\alpha$, $\varphi$, $\zeta$, and $\beta$, and the observed 
spectral luminosity $L_E$ varies periodically with $\varphi$. The magnitude 
of the variation may be described by the ``pulsed fraction,''
\beq
\label{eq:PF}
   f_p(E)\equiv \frac{L_E^{\max}-L_E^{\min}}{L_E^{\max}+L_E^{\min}},
\eeq
or the ``pulsed spectrum,''
\beq
\label{eq:Lpuls}
   L_E^{\rm puls}\equiv L_E^{\max}-L_E^{\min}.
\eeq
If the emission is not resolved in rotational phase, its observed spectrum is 
obtained by averaging the instantaneous $L_E$ over the rotation period,
\beq
\label{eq:Lav}
    \bar{L}_E=\frac{1}{2\pi}\int_0^{2\pi} L_E\, d\varphi.
\eeq
The result depends on $\alpha$, $\beta$, and $\zeta$. If the emitting 
region (j-bundle) is symmetric about the magnetic axis, the result only 
depends on $\alpha$ and $\beta$.

\subsection{Aligned rotator}

Aligned rotator has $\alpha=0$ (magnetic axis parallel to the rotation axis).
Then $\varphi=\phi$ and averaging over rotation is the same as averaging
over the magnetic azimuthal angle $\phi$.
The shape of the averaged spectrum $\bar{L}_E$ does not depend on the 
azimuthal extension of the j-bundle $\phil$; e.g. one may chose
$\phil=2\pi$ (axisymmetric j-bundle). 

The spectrum produced by the axisymmetric j-bundle is shown in Figure~7.  
It strongly depends on the angle between the line of sight and the magnetic 
axis. In particular, the position of the high-energy peak significantly varies 
around 1~MeV. One can qualitatively understand the variations in 
the spectrum with changing
inclination by combining the following facts.
(1) Emission from the loop is generated on field lines that have footpoints at 
co-latitudes $\theta\sim 0.3$ and reach the equatorial plane at radii 
$\Rmax=10-20$.
(2) Emission is strongly beamed outward along the magnetic field lines. 
(3) Photons of a given energy $E$ are mainly generated where 
$b\approx 0.1 (E/m_ec^2)^{1/2}$ (see \Eq~\ref{eq:E}).
Note also that observers in the equatorial plane (inclination 
$\theta=90^{\rm o}$) are special --- by symmetry, both hemispheres equally 
contribute to the observed spectrum.
At small inclinations, one hemisphere strongly dominates hard 
X-ray emission; however, the contribution of the other hemisphere is still 
significant in the soft-X-ray band.

\begin{figure}
\epsscale{1.1}
\plotone{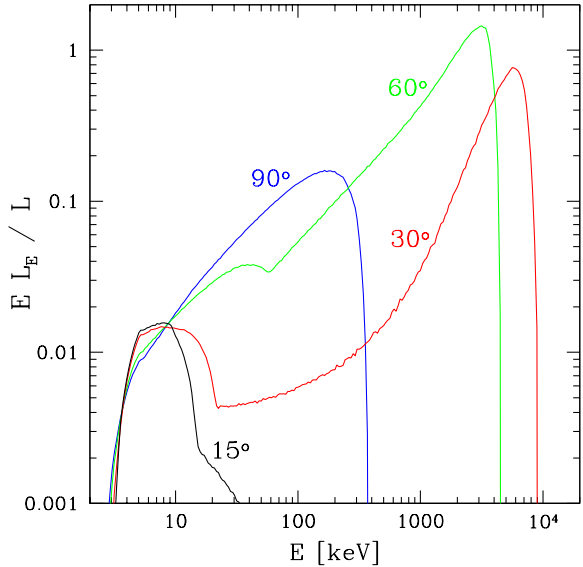}
\caption{Spectrum from the axisymmetric j-bundle, observed at four
different angles with respect  to the magnetic axis: $15^{\rm o}$, 
$30^{\rm o}$, $60^{\rm o}$, and $90^{\rm o}$.
The spectrum is normalized to the total luminosity of the j-bundle $L$ given 
by \Eq~(\ref{eq:Ltot}).
}
\end{figure}
\begin{figure}
\epsscale{1.25}
\plotone{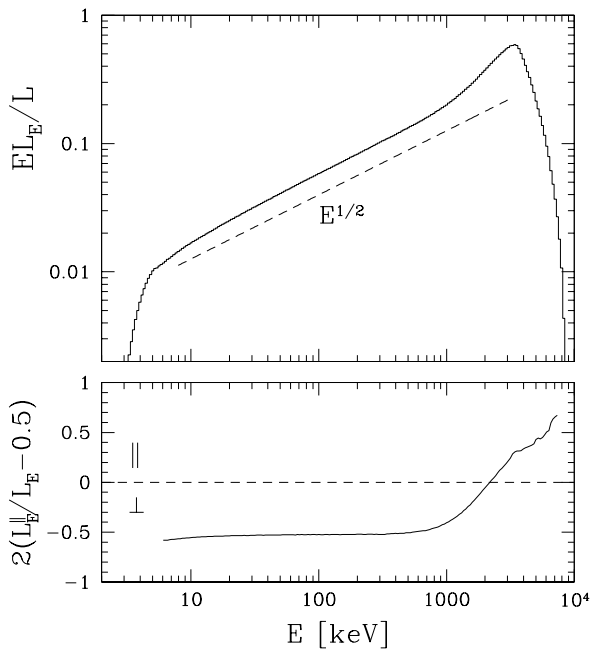}
\caption{Upper panel: spectrum from the axisymmetric j-bundle, $L_E$, 
averaged over the object inclination to the line of sight. 
Lower panel: relative contributions of $\perp$ and $\parallel$ to the 
total spectrum $L_E=L_E^\perp+L_E^\parallel$.
}
\end{figure}

If the emission shown in Figure~7 is averaged over cosine of inclination angle
one obtains the spectrum shown in Figure~8. 
Below 1~MeV, it approximately follows the 
power-law $EL_E\propto E^{1/2}$ in agreement with \Eq~(\ref{eq:LE})  
derived in \Sect~3.4. It corresponds to photon index $\Gamma=1.5$
(which is defined by $L_E\propto E^{-\Gamma+1}$).
This does not imply that, given a random inclination, the most probable 
observed spectrum has $\Gamma= 1.5$.  Significantly harder slopes 
$\Gamma\sim 1$ are observed in Figure~7.
In a range of inclinations, the spectra have breaks well below 1~MeV, and 
after averaging over inclination this gives the featureless power law with a 
softer slope $\Gamma=1.5$.

Approximately 3/4 of photons scattered in the j-bundle have 
the $\perp$ polarization.
Therefore at energies $E<1$~MeV (where the splitting of $\perp$ 
photons is negligible), the produced emission is dominated by the $\perp$
polarization, see the lower panel in Figure~8.

The inclination-averaged spectrum exhibits a significant 
hump at $E\simgt 1$~MeV. It is the result of photon splitting.
The daughter photons produced by splitting have the $\parallel$ polarization,
and the splitting-dominated  
component is particularly visible in the lower panel of Figure~8
that shows the polarization of emitted photons. Note that 
Figure~8 does not take into account the possible multiple scattering of 
high-energy photons on their way out of the j-bundle (\Sect~3.5). 
 Additional scattering weakly changes
the spectrum of escaping radiation, however it
can significantly affect polarization of the MeV hump. 
Scattering of daughter photons after splitting switches polarization from 
$\parallel$ to $\perp$ with probability of $3/4$ and tends to 
bring the polarization to the same value as below 1~MeV.

No pulsations are produced by the aligned rotator if the j-bundle is 
axially symmetric, $\phil=2\pi$. Strong pulsations
are produced if $\phil<2\pi$, i.e. if the j-bundle is confined to a smaller 
range of azimuthal angles. The pulsations are the consequence of 
strong beaming of high-energy emission along the magnetic 
field lines, which makes the emission almost invisible when the line 
of sight is outside the interval $0<\phi<\phil$. The pulse width is proportional
to $\phil$, and the pulse fraction can approach 100\%.

\subsection{Orthogonal rotator}

Orthogonal rotator has $\alpha=\pi/2$  (magnetic axis is perpendicular
to the rotation axis). Then even an axisymmetric j-bundle
($\phil=2\pi$) will produce pulsating emission. As the magnetar rotates,
the inclination of line of sight to the magnetic axis, $\theta$, 
periodically changes and the observer samples 
the spectra shown in Figure~7.
The spectrum strongly depends on $\theta$ and hence significant periodic 
variations in $L_E$ are expected, depending on $\beta$ (the angle between
the line of sight and the rotation axis).
In particular, the position of the high-energy peak will oscillate with a 
significant amplitude. No pulsations are expected only if $\beta=0$; 
the observed spectrum in this case is the same as the $\theta=\pi/2$ 
spectrum in Figure~7.

Figure~9 shows the emission from the orthogonal rotator viewed at 
$\beta=20^{\rm o}$. The rotation-averaged spectrum has the photon 
index $\Gamma\approx 1$ at photon energies below 70~keV. 
The pulsed spectrum is significantly harder, $\Gamma\approx 0$. 
The pulsed fraction increases with 
$E$ and approaches 100\% at $E\approx 300$~keV.
The model is close to observational data reported for 4U~0142+61 and 
1RXS J170849-400910 --- two magnetars whose hard X-ray pulsations
have been studied in detail (den Hartog et al. 2008a,b).

Figure~10 shows the emission from the orthogonal rotator viewed at 
$\beta=90^{\rm o}$.
The rotation-averaged spectrum has the photon index $\Gamma\approx 1.5$
below 1~MeV; this spectrum is close to the average spectrum in Figure~8, 
although not exactly the same.\footnote{
     For the orthogonal rotator viewed at $\beta=90^{\rm o}$, we 
     have $\theta=\varphi$, where $\theta$ is the angle between the 
     line of sight and the magnetic axis. Therefore, averaging over rotation 
     $0<\varphi<2\pi$ is equivalent to averaging over $\theta$.
     In Figure~8, the emitted spectrum was averaged over $\cos\theta$
     (rather than $\theta$), giving a slightly different result.}
The pulsed spectrum is again harder than the average
spectrum. The amplitude of pulsations is significantly larger than in 
Figure~9; the pulsed fraction approaches 100\% at $E\approx 20$~keV.

\begin{figure}
\epsscale{1.1}
\plotone{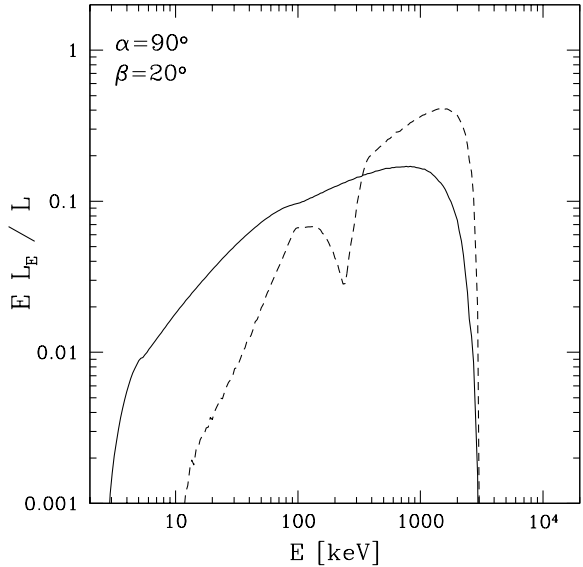}
\caption{Spectrum from the orthogonal rotator ($\alpha=90^{\rm o}$)
when the line of sight is at angle $\beta=20^{\rm o}$ with respect to 
the spin axis. The j-bundle is assumed to be symmetric about the magnetic 
axis ($\phil=2\pi$). Solid curve shows the spectrum averaged over 
rotation (\Eq~\ref{eq:Lav}). 
Dashed curve shows the pulsed spectrum (\Eq~\ref{eq:Lpuls}).
}
\end{figure}
\begin{figure}
\epsscale{1.1}
\plotone{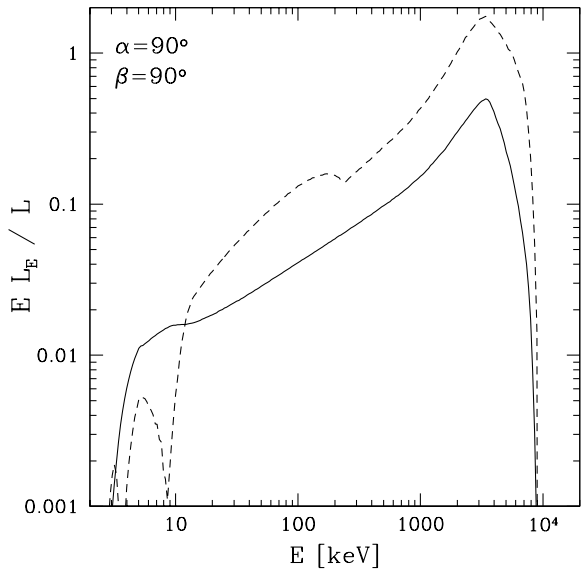}
\caption{Same as Figure~9 but for $\beta=90^{\rm o}$.
}
\end{figure}

The variety of observed spectra and pulse profiles becomes even larger 
if the j-bundle is not axisymmetric.
Then the observed emission depends on the third Euler angle $\zeta$ 
(\Sect~4.1). The pulse profile in a given spectral window depends
on the angles $\alpha$, $\beta$, $\zeta$ and the geometry of the j-bundle.


\section{Discussion}
\label{sec:disc}

In this paper, we calculated the X-ray emission expected from 
large twisted magnetic loops around magnetars.
The loops are inevitably heavily loaded  with $e^\pm$ pairs, 
with multiplicity $\M\sim 10^2$, which leads to efficient screening of electric fields.
The created plasma outflows along the magnetic loop and loses its energy 
to radiation.
Transformation of the outflow energy to escaping radiation occurs in 
the region $10^{13}> B>3\times 10^{11}$~G, which requires a minimum 
size of the loop 3 to 10 stellar radii (if the surface field is $\sim 10^{15}$~G).
We did not consider small loops because most of photons scattered near 
the star are unable to escape the magnetosphere; besides, smaller loops
are less likely to be active (\Sect~2.1).

A key feature of the $e^\pm$ outflow along the loop is that its Lorentz factor is 
self-regulated so that it scatters radiation with $\gsc\approx 100 B/\BQ$ 
(\Sect~2.2). This fact determines the radiation spectrum emitted by the outflow,
taking into account that the scattered photons have the characteristic energy
$E\sim \gsc (B/\BQ)m_ec^2$. The emitted spectrum below 1 MeV is hard, i.e.
softer photons $E\ll 1$~MeV contribute less to the outflow luminosity. 
This is simply because softer photons are emitted where the outflow 
decelerates, and there remains less energy to emit.
The spectrum is also suppressed at energies $E\gg 1$~MeV, because photons 
of such high energies are emitted in the zone of strong $B$ and cannot escape. 

The model predictions appear to agree with available data.
Observations of magnetars in the 10-100~keV band show 
a typical photon index $\Gamma\sim 1-1.5$,
similar to what is found in the model spectra. 
The observed spectra peak above 100~keV, and the existing upper limits for 
AXP 4U~0142+61 at 1-30~MeV indicate a spectral break between 
0.3 and 1~MeV (den Hartog et al. 2008a). 

The strong beaming of radiation along the loop explains why the observed
hard X-ray flux shows huge pulsations, with the pulsed fraction growing with 
photon energy up to 100\% (Kuiper et al. 2003; den Hartog et al. 2008a,b). 
No unique pulse profile is predicted by our model, because the pulse is
sensitive to the loop geometry, its orientation with respect to the rotation 
axis and the line of sight. The pulse profile may be further complicated 
if more than one loop are activated and the superposition of their emissions
is observed. The expected hard X-ray pulse does not have to be in phase 
with the soft X-ray pulse, as observed
in AXP 1RXS~J170849-400910 (see Figure~6 in Den Hartog et al. 2008b).

In general, sources of beamed, pulsed radiation are not expected to display a 
standard spectrum at all viewing angles, and magnetars are not an exception.
The active loop typically emits a hard spectrum, $\Gamma\sim 1-3/2$,
but its detailed shape depends on the loop and the viewing angle
(Figures~7, 9, 10).
The predicted high-energy peak significantly depends on the line of 
sight and its observed position is expected to oscillate as the magnetar rotates.

Note also that a very hard spectrum $\Gamma<1$ may be produced if 
smaller magnetic loops are activated. For example, consider a loop that 
is confined to the region of strong magnetic field $B>3\times 10^{12}$~G,  
where $\hbar\omega_B>35$~keV.
The loop generates photons of energy $E\sim \gamma\hbar\omega_B$ 
and almost no emission is produced below 35~keV, leading to a hard 
spectral slope.

The surface transition layer between the corona and the star 
may give a second contribution to the hard X-ray emission
(Thompson \& Beloborodov 2005; Beloborodov \& Thompson 2007). 
This dense layer is bombarded by the particles from the discharge
and may reach temperatures $kT\simgt 100$~keV, sufficient to emit 
high-energy bremsstrahlung radiation with photon index $\Gamma\approx 1$.
Observed magnetars show complex spectral variations with rotational phase, 
which suggests that the emitting plasma occupies an extended region 
in the magnetosphere rather than a heated spot on the surface.
This favors the outflow model described in the present paper.
Polarization measurements would provide a further
test discriminating between the two mechanisms.
The surface transition layer emits photons with the $\parallel$ polarization,
while the loop itself emits radiation that is dominated by the $\perp$ 
polarization (also called ``extraordinary'' or E-mode).
Careful calculations of polarization should include transfer 
effects (Fern\'andez \& Davis 2011).

\subsection{Annihilation line}

The outflowing $e^\pm$ pairs accumulate and annihilate at the top of the 
magnetic loop with Lorentz factors $\gamma\sim 1$ (Figure~1). This 
must create an annihilation line at $E\approx 0.511$~MeV, which we 
did not show in our sample spectra. There is a 
simple relation between the luminosity of annihilation radiation $\Lann$ and
the total outflow power $L$ (which is equal to the luminosity of the hard 
X-ray component). 
The outflowing particles enter the radiative zone  $B<10^{13}$~G
with the average Lorentz factor $\gav_0\sim 10-20$
and radiate almost all their kinetic energy $(\gav_0-1)m_ec^2$
in hard X-rays before annihilating. Therefore, a fraction of $\sim\gav_0^{-1}$
of the total outflow energy is converted to annihilation radiation, 
\beq
\label{eq:Lann}
    \frac{\Lann}{L}=\frac{1}{\gav_0}\sim 0.1.
\eeq
The visibility of the annihilation line at 511~keV depends on its width.
As $e^\pm$ are decelerated before annihilating, a pronounced
narrow line is expected. Mildly relativistic motions
in the annihilation region would make the line broad and less visible. 
Note also that $\Lann$ is approximately isotropic, while $L$ is strongly anisotropic;
therefore the observed ratio $\Lann/L$ will depend on the line of sight and 
can be significantly larger or smaller than the mean expectation given by 
\Eq~(\ref{eq:Lann}).

\subsection{Temporal behavior of activity}

The nonthermal emission of magnetars must be fed by 
ohmic dissipation of electric currents excited in their twisted
magnetospheres (\Sect~2.1). 
The rate of energy release is proportional to voltage established 
along the magnetic field lines, which is regulated by electric discharge 
to $\Phi\sim 10^9$~V, in agreement with the observed luminosity.
One expects the untwisting magnetosphere and its luminosity to evolve 
on the ohmic timescale, which is comparable to 1~year.
Several transient magnetars show evidence for this evolution
(Beloborodov 2011); XTE~J1810-197 is a canonical example.
Hard X-ray data for transient magnetars are meager.
A strong evolution of the hard X-ray flux is expected in these objects, which
may be tested by the upcoming {\it NuSTAR} observations.

Other sources, in particular 4U~0142+61 and 1RXS J170849-400910, 
remain stable over a long observation period (about 10~years).
Their hard X-ray emission has been studied in detail (den Hartog et al. 
2008a,b).
The different patterns of magnetar activity are not understood because of
our ignorance  concerning the pattern and speed of the crustal motions 
that cause the activity. The crust is likely to 
flow plastically and gradually twist the magnetosphere.
It is unclear whether a quasi-steady state should be expected, punctuated by 
sudden outbursts triggered by excessive twisting (Parfrey et al. 2012b).

\subsection{1-10~keV emission}

In this paper, we focused on the mechanism for the hard X-ray peak 
in magnetar spectra, and did not address the shape of the soft X-ray 
(keV) peak. As has been known for a long time, the keV peak 
is different from simple blackbody (e.g. Woods \& Thompson 2006). 
The 1-10~keV spectrum is explained as a 
superposition of the thermal component and a soft power-law tail with 
photon index $\Gamma=2-4$. The soft tail is distinct from the 
hard X-ray component, although a correlation between them
was reported (Kaspi \& Boydstun 2010).

Previous models of resonant scattering around magnetars were able 
to reproduce the soft tail in the 1-10~keV band (Fern\'andez \& Thompson
2007; Nobili, Turolla, \& Zane 2008; Rea et al. 2008). In these models,
the magnetospheric region where $\hbar\omega_B\sim 1-10$~keV
($B=10^{11}-10^{12}$~G) is filled with a mildly relativistic plasma 
($\gamma\beta\sim 1$). It is usually assumed that the
plasma is made of counter-streaming positive and negative charges,
which maintain the electric current with particle multiplicity $\M\sim 1$;
then the scattering optical depth 
can be estimated as $\tau\sim \psi/\beta\sim 1$ (Thompson et al. 2002).
The results of the present paper show that
these assumptions are problematic. 
The typical pair multiplicity $\M$ must be comparable to $10^2$, and 
charges of both signs should stream away from the star, toward their 
annihilation at the top of the magnetic loop. The model of the soft X-ray 
tail due to resonant scattering should be revised accordingly.

The high multiplicity alleviates the following problem.
The particle flux along the magnetic field lines 
is $\dNe=\M I/e$ where $I/e\sim 5\times10^{37}\,\psi\,{\rm s}^{-1}$ 
is a typical electric current circulating through the region 
$\hbar\omega_B\sim 3$~keV (cf. \Eq~\ref{eq:I1}).  
The observed luminosity in the soft tail $\Ltail$ is comparable to 
$10^{35}$~erg~s$^{-1}$, and hence the energy emitted per particle is given by 
\beq
\label{eq:Ls}
     \frac{\Ltail}{2\dNe}=\frac{e\,\Ltail}{2I\,\M} \sim 600 \, \M^{-1}
        \left(\frac{\Ltail}{10^{35}~{\rm erg~s}^{-1}}\right) {\rm ~MeV}.
\eeq
The mildly relativistic particles that are supposed to produce $\Ltail$ 
have energies $\sim 1$~MeV. Hence the picture of 
counter-streaming charges with $\M\sim 1$ requires
that the emitted energy is 2-3 orders of magnitude larger than the 
particle energy.\footnote{One could assume that a voltage $e\Phi\sim 1$~GeV
       is applied to this region and responsible for the enhanced emission rate per
       particle. In this picture, radiative drag impedes the electrostatic 
       acceleration and keeps the particles mildly relativistic. This ``radiatively locked 
       flow'' is considered in more detail in the accompanying paper (B12).
       Note also that the balance between the radiative
       and electrostatic forces is unstable when the 
       particle interacts with photons below the Wien peak of thermal radiation.} 
In contrast, the model with multiplicity $\M\simgt 10^2$ implies that a few 
MeV per particle is sufficient to feed the observed $\Ltail$ and modify the thermal peak.

\acknowledgments
This work was supported by NASA NNX10AI72G.


\begin{appendix}

\section{Resonant scattering}

\subsection{Cross section, photon energy, and polarization}

The polarization states of X-rays in the magnetosphere are determined by the 
quantum electrodynamic effect of vacuum polarization in a strong magnetic field.
It defines two eigen modes for an electromagnetic wave with a 
wavevector $\bk$:  with electric field oscillating along $\bB\times\bk$ 
($\perp$ polarization, sometimes called ``extraordinary'' or E-mode)
and with electric field oscillating along $\bk\times(\bB\times\bk)$ 
($\parallel$ polarization, sometimes called ``ordinary'' or O-mode).

In our problem, the outflow is ultra-relativistic in the region of strong $B$,
and target photons are strongly aberrated to become nearly 
parallel to $\bB$ in the electron rest frame. Then the scattering
cross section is the same for $\parallel$ and $\perp$ photons.
Let $\Ec_t=\hbar\tilde{\omega}$ be the energy of a target photon 
measured in the electron rest frame. 
Resonant scattering cross section in this frame is given by
\beq
\label{eq:sigma}
    \tilde{\sigma}=2\pi^2r_ec\, \delta(\omegac-\omega_B), \qquad 
    \omega_B=\frac{eB}{m_ec}, 
\eeq
where $r_e=e^2/m_ec^2$.
\Eq~(\ref{eq:sigma}) can be derived classically when $B\ll\BQ$
(e.g. Canuto et al. 1971; Ventura 1979). 
It also remains valid at any $B$,
as long as target photons propagate nearly along $\bB$ in the electron 
rest frame (e.g. Harding \& Daugherty 1991).

Consider an electron moving along a magnetic field line with Lorentz factor
$\gamma=(1-\beta^2)^{-1/2}$. When the electron scatters a photon,
the photon energy changes from $E_t$ to $E$.
Energy conservation gives $(\gamma-\gamma_f)\, m_ec^2=E-E_t$ 
where $\gamma_f$ is the electron Lorentz factor after scattering.
Resonant scattering may be viewed as photon absorption, which excites the
electron to the first Landau level $E_B=(1+2b)^{1/2}m_ec^2$, followed by
photon emission (electron de-excitation).  The Lorentz factor of the excited electron $\gamma_1$ is found from $\gamma\,m_ec^2+E_t=\gamma_1 E_B$. 
Assuming that de-excitation on average does not decelerate or accelerate
the electron along $\bB$, the mean expectation for $\gamma_f$ is 
$\gamma_1$. Then the average energy of emitted photons is 
$E=(\gamma m_ec^2+E_t)(1-m_ec^2/E_B)$, which yields (using 
$E_t\ll \gamma m_ec^2$),
\beq
\label{eq:Eav}
    E\approx\gamma\,m_ec^2 \left(1-\frac{1}{\sqrt{1+2b}}\right),
      \qquad b=\frac{B}{\BQ}.
\eeq
This expression gives a typical energy of scattered photons;
the exact $E$ in each scattering event depends on the scattering angle. 

The scattered photons can be in one of the two polarization states. 
Using the accurate relativistic differential cross section,
we find that the fraction of scattered photons with the 
$\perp$ polarization is conveniently approximated by the following formula,
\beq
\label{eq:polariz}
   \frac{\sigma_\perp}{\sigma_\perp+\sigma_\parallel}
   \approx \frac{3+b/2}{4+b}.
\eeq

\subsection{Scattering rate}

Let $n_\omega(\vec{\Omega},\omega)$ be the local angular distribution of 
the photon spectral density [cm$^{-3}$~s]. The cross section (\ref{eq:sigma})
implies the following rate of resonant scattering by one electron,
\beq
\label{eq:dNsc}
   \dot{N}_{\rm sc}=\frac{2\pi^2 r_e c^2}{\gamma}
       \int d\Omega \int d\omega\, n_\omega
               \,\delta(\omega-\omega_{\rm res}),
\eeq
where 
$\omega_{\rm res}=\gamma(1-\vec{\beta}\cdot\vec{\Omega})\,\omega_B$.
In our problem, the spectrum of target radiation seen by the electron may be 
approximated by a (diluted) Planckian with temperature $T$ and photon density $n$,
\beq
        \omega\,n_\omega\approx 0.42\,\frac{y^3}{e^y-1}\,n\, 
             f(\vec{\Omega}),
                  \qquad y=\frac{\hbar\omega}{kT}.
\eeq
Then $\dot{N}_{\rm sc}$ becomes,
\beq
\label{eq:y}
   \dot{N}_{\rm sc}\approx \frac{2\pi^2 r_e c^2\,\hbar}{\gamma\, kT}\,0.42\,n\,
          \int f(\vec{\Omega})\, \frac{\yres^2}{\exp \yres-1}\, d\Omega.
\eeq
The resonance condition defines the range of possible $\yres$, 
$\ymin<\yres<\ymax$.  When $\ymin\gg 1$, only photons in the Wien tail 
of the Planck spectrum can be scattered. Then the integral in 
\Eq~(\ref{eq:y}) strongly peaks where $\yres\approx\ymin$, which gives
(using $f[\vec{\Omega}]d\Omega\sim \yres^{-1}d\yres$),
\beq
\label{eq:dNsc1}
   \dot{N}_{\rm sc}\approx \frac{2\pi^2 r_e c^2\,\hbar}{\gamma kT}\,0.42\,n\,
       \ymin e^{-\ymin},
 \qquad  \ymin=\frac{\hbar\omega_B}{\gamma(1-\beta\cos\ang_{\max})\,kT},
\eeq
where $\ymin$ corresponds to the largest photon angle $\ang_{\rm max}$ with 
respect to $\vec{\beta}$. Thermal photons reflected by opaque plasma near
the top of magnetic loops (Figure~1) have the maximum angle, as they stream 
against the outflow direction.
The exponential sensitivity of  $\dot{N}_{\rm sc}$ to $\ang_{\rm max}$ explains 
why the reflected photons are 
the main targets for scattering even though their total number density $n$ 
is much smaller than that of  photons streaming directly from the star.
For photons streaming at smaller angles $\ang$, $\ymin$ 
would increase, leading to an exponential suppression of $\dot{N}_{\rm sc}$. 
For instance, changing $\ymin=7$ to $\ymin=14$ reduces 
$\dot{N}_{\rm sc}$ by the factor of $4\exp(-7)\approx 4\times 10^{-3}$.

The density of reflected thermal radiation that should be used in 
\Eq~(\ref{eq:dNsc1}) is given by
\beq
      n\sim \frac{\xi\,\Lstar}{4\pi R_1^2\, 2.7kT\, c}
        \approx 10^{17}\,
          \left(\frac{\xi}{0.1}\right)
          \left(\frac{\Lstar}{10^{35}{\rm ~erg~s}^{-1}}\right)
          \left(\frac{kT}{0.5 \rm ~keV}\right)^{-1} 
          \left(\frac{R_1}{100 \rm ~km}\right)^{-2} 
               {\rm ~cm}^{-3}.
\eeq
Here $\Lstar$ is the thermal luminosity of the star, $2.7kT\sim 1$~keV stands 
for the average energy of thermal photons, and $\xi$ is the fraction of $\Lstar$ 
that is intercepted by the slow, opaque plasma at the top of magnetic loops.

In the calculations in this paper, it is convenient to describe the number of 
thermal photons scattered by an outflowing particle
as a function of decreasing magnetic field, 
\beq
  \frac{dN_{\rm sc}}{dB}=\frac{1}{\beta c}\,\frac{\dot{N}_{\rm sc}}{dB/dl},
\eeq
where $l$ is length measured along the magnetic field line, and 
$dB/dl$ (for dipole field lines) is given in \Eq~(\ref{eq:dbdl});
in the simulations shown in Figures~2 and 3 we use a simplified $dB/dl=-3B/r$.

\subsection{Drag exerted by the reflected radiation on a relativistic electron}

The electron loses energy to radiation with rate
\beq
   \dot{E}_e\approx -\dot{N}_{\rm sc}\,(E-E_t),
\eeq
where $E$ is the average energy of scattered photons (\Eq~\ref{eq:Eav})
and $E_t$ may be neglected compared with $E$. This gives
\beq
   \frac{\dot{E}_e}{m_ec^3}\approx - \frac{n\,\sT}{\alf}\,\frac{m_ec^2}{kT}
        \, \ymin\, e^{-\ymin}\left(1-\frac{1}{\sqrt{1+2b}}\right).
\eeq
where $\alf=e^2/\hbar c=1/137$. The corresponding drag force acting on the 
relativistic electron is given by $\F\approx\dot{E}_e/c$. 

The expression for $\dot{E}_e$ may be written in a different, physically
more transparent form,
\beq
\label{eq:F}
   \dot{E}_e\approx - c\,\sigres\,\ntarget\, E.
\eeq
Here $\sigres$ is the effective cross section for resonant scattering,
\beq
\label{eq:sigres}
    \sigres\approx (1-\beta\cos\ang)\,\pi r_e\tilde{\lambda},
\eeq
where $\tilde{\lambda}=2\pi c/\omega_B$ is the photon wavelength in the 
rest frame of the electron. 
The factor $(1-\beta\cos\ang)$ (from transformation of cross section to the 
lab frame) is comparable to unity.
$\ntarget$ in \Eq~(\ref{eq:F}) is the density of photons near 
the resonant energy $\hbar\omega_{\rm res}$.
Our electron interacts with photons in the Wien tail of the spectrum, and
their number peaks near the minimum $\yres=\hbar\omega_{\rm res}/kT$,
\beq
\label{eq:nt}
   \ntarget\approx kT\left.\frac{dn}{dE_t}\right|_{\ymin}
   \approx 0.42\, \ymin^2 e^{-\ymin}\,n.
\eeq
Substitution of \Eqs~(\ref{eq:sigres}) and (\ref{eq:nt}) to \Eq~(\ref{eq:F}) gives 
an explicit expression for the drag force $\F=\dot{E}_e/c$.
The estimate for $\F$ may be summarized in the following form, 
\beq
\label{eq:F2}
    \frac{r\,\F}{\gamma\,m_ec^2}\approx
    \frac{3\pi}{4}\,\frac{\sT n_t r}{\alf}\,q\,(1-\beta\cos\ang_{\max}) 
          \sim - 10\,\frac{r}{R}\,q\,n_{17}\,\ymin^2\,e^{-\ymin},
\eeq
where $n_{17}=n/10^{17} {\rm ~cm}^{-3}$.
The recoil correction factor $q(b)=1$ when $b\ll 1$ (\Eq~\ref{eq:Esc}).
The approximate estimate given in \Eq~(\ref{eq:F2}) is sufficient in view of 
the strong exponential dependence on $\ymin$ and the large prefactor 
$10(r/R)\,n_{17}\, \ymin^2$. The outflowing particles surf with 
$r\F/\gamma m_ec^2\sim |d\ln B/d\ln r|\sim 3$ (\Sect~2.2), 
which corresponds to $\ymin\sim 7$ and the prefactor $\sim 10^3$.


\section{Photon splitting and conversion to electron-positron pairs}

\subsection{Conversion to $e^\pm$}

Consider a photon of energy $E$ propagating at an angle $\ang$ with respect
to the magnetic field lines. The photon can convert an $e^\pm$ pair if the 
threshold condition $\Ethr<E$ is satisfied. The threshold energy depends on 
the photon polarization: $\Ethr=2m_ec^2/\sin\ang$ for $\parallel$ photons
(then both $e^+$ and $e^-$ can be created in the ground Landau state) 
and $\Ethr=2m_ec^2(1+2b)^{1/2}/\sin\ang$ for $\perp$ photons (then at 
least one of the created particles is required to be in an excited Landau state).

The rate of conversion depends on the local magnetic field $b=B/\BQ$. 
If $b\simgt 0.05$, one can assume practically instantaneous conversion 
once the thershold condition is satisfied; then the $e^\pm$ pair is created
with the lowest possible energy $E_-+E_+=\Ethr$. 
If $b\simlt 0.05$, conversion is significantly delayed and generally occurs 
when $\Ethr$ is significantly below $E$. Then $e^\pm$ are created in high 
Landau states, and the absorption coefficient for the photon is given by 
$ \kappa_{\rm abs}\approx 4.3\times10^7\,b\,
            \exp\left(-8m_ec^2/3\,b\,E\sin\ang\right){\rm ~cm}^{-1}$ (Erber 1966).

The model described in this paper
predicts that practically all photons that convert to $e^\pm$ do so in the region 
$b>0.1$, and very close to their emission points. Then the photon energy 
$E$ is shared by two particles created in
the ground Landau state with energies $E/2$.

\subsection{Splitting}

Following selection rules (Adler 1971; Usov 2002), 
we assume that only one splitting channel is allowed: 
$\perp\rightarrow \parallel + \parallel$
(Adler [1971] uses the opposite notation for the polarizations states, 
$\perp\leftrightarrow\parallel$).
The absorption coefficient for a $\perp$ photon of energy $E$ 
propagating at an angle $\ang$ with respect to the magnetic field lines
is given by
\beq
     \kappa_{\rm sp}=0.12\,\zeta\,(b\sin\ang)^6\,\left(\frac{E}{m_ec^2}\right)^5
       {\rm ~cm}^{-1},
\eeq
where $\zeta=1$ if $b\ll 1$ and 
$\zeta\sim 4b^{-6}$ in the limit of $b\gg 1$ 
(the exact asymptotic value depends on $E\sin\ang$, see Baring \& Harding 1997).
Factor $\zeta$ can be calculated numerically (Figure~8 in Adler [1971] 
and Figure~1 in Baring \& Harding [1997]). Medin \& Lai (2010) and our
Monte-Carlo simulations use an approximate formula,
\beq
    \zeta=(g+0.05)^{-1}(0.25g+20)^{-1}, \qquad 
        g=b^3\exp\left[-0.6\left(\frac{E\sin\ang}{2m_ec^2}\right)^3\right].
\eeq 
When a photon splits, its energy $E$ is shared between two daughter 
photons propagating in the same direction. 
The probability distribution for the energy a daughter photon $E_1$
is approximately given by $p(u)\approx 30\,u^2(1-u)^2$, where 
$u=E_1/E$ (Baring \& Harding 1997). We use this distribution
in our Monte-Carlo simulations.

\subsection{Implementation of splitting and conversion in Monte-Carlo 
simulations}

To determine the fate of high-energy photons generated by resonant 
scattering we follow the 
photon along its trajectory and track the change of its pitch angle $\ang$ 
with respect to the local magnetic field lines. The (initially small) angle 
grows due to the curvature of the field lines while
the photon polarization remains unchanged along the trajectory 
(the small difference of the refraction indices for the two polarization states 
results in locking/adiabatic tracking of the photon polarization).
For photons with $\parallel$ polarization,
we check if they convert to $e^\pm$. The threshold for conversion 
$\Ethr=2m_ec^2/\sin\ang$ is reduced as $\ang$ increases.
When $\Ethr/E$ is reduced below unity, conversion 
becomes possible, and in practice occurs immediately.
For photons with $\perp$ polarization, we check if they split or convert
to $e^\pm$. In practice conversion is negligible as splitting occurs before
the conversion threshold condition is satisfied.

Consider a photon emitted along the outflow direction at point $\rem$.
Let $\Emax(\rem)$ be the maximum energy for the photon to avoid 
destruction and escape. More precisely, we define 
two maximum energies $\Emax^\parallel(\rem)$
and $\Emax^\perp(\rem)$ for $\parallel$ and $\perp$ photons, respectively.
For any given magnetospheric configuration, $\Emax^\parallel(\rem)$
and $\Emax^\perp(\rem)$ can be calculated numerically by 
following the photon trajectories as described above. 

The fate of an emitted photon in our Monte-Carlo simulations is determined
by comparing its energy $E$ with $\Emax$. The simulation can be 
further simplified due to the following observation: 
if a resonantly scattered photon is destroyed, the 
destruction happens quickly, close to the emission point $\rem$
(except for a negligible number of photons scattered near the magnetic axis,
where the field line curvature is small).
This is because the high-energy photons capable of conversion or 
splitting are generated in the region of strong and curved magnetic field 
$b\simgt 0.1$
(recall that the energy of a resonantly scattered photon does not exceed 
$2\gsc b\,m_ec^2\approx 200 b^2\,m_ec^2$).
In the Monte-Carlo simulations it is a good approximation to assume the
immediate conversion of $\parallel$ photons if $E>\Emax^\parallel(\rem)$ 
and the immediate splitting of $\perp$ photons if $E>\Emax^\perp(\rem)$. 
This approximation is used in all sample models shown in \Sect~4. 

Let $\bar{E}(\rem)$ be the average energy of resonantly scattered photons.
Most of the photons generated in the zone $\bar{E}>\Emax$ are destroyed.
As the outflow moves along a field line,
there is a sharp transition to photon escape at the boundary $\rb$ defined 
by $\bar{E}(\rb)=\Emax(\rb)$.
Similar results will be obtained if one assumes that all photons scattered 
with $E<\Emax(\rb)$ escape (regardless of their emission points $\rem$),
and all photons with $E>\Emax(\rb)$ split or convert, 
depending on their polarization states.
Such an approximation is used in the cascade simulations in \Sect~3.2.
Instead of the pre-calculated $\Emax^\parallel(\rem)$ and 
$\Emax^\perp(\rem)$ (which depend on the chosen dipole magnetic 
configuration), we assume a given fixed 
$\Emax^\parallel=\Emax^\perp\approx \gsc(b_0)\,b_0\,m_ec^2$.
The value of $\Emax$ is typically somewhat above 1~MeV,
depending on the curvature of magnetic field lines.
Simulations shown in Figures~3-6 assume $E_{\max}=3m_ec^2$ --- 
a typical value obtained with the more detailed simulations that follow the 
propagation of photons through the magnetosphere.

\subsection{Boundary between the pair-loading and radiative zones}

The location of the boundary between the two zones was estimated
in \Sect~2.3. Approximately, it is where the magnetic field $b=B/\BQ$ 
equals $b_0\sim 1/4$. The exact $b_0$ varies for different field lines of the 
magnetosphere. It is convenient to label the field lines by their 
maximum extension radius $\Rmax$ (the apex radius).
The boundary location also depends on the magnetic moment of the star 
$\mum$ and its temperature $T$. 
These dependencies may be formulated once we specify a more rigorous
definition of the boundary.

To define the radiative zone, consider a photon scattered at a given point P by an 
electron with Lorentz factor $\gsc$ (\Eq~\ref{eq:gsc1}). We focus on the region where 
$b\ll 1$ and $\gsc\gg 1$, as the estimated zone boundary is in this region. 
The energy of the scattered photon is given by 
$E=\gsc(1+\beta_{\rm sc}\cos\angc)\,b\,m_ec^2$
where $\angc$ is the scattering angle in the electron rest frame. The mean 
expectation for $\cos\angc$ is zero and the average photon energy after scattering 
is $\gsc\,b\,m_ec^2$. The photon direction in the lab frame is nearly aligned with 
the electron velocity, within an angle $\ang\approx \gsc^{-1}$, and hence the photon 
is emitted nearly parallel to the local magnetic field.
The photon propagates along the straight line through the magnetosphere
(the bending of photon trajectory by neutron-star gravity is neglected).
We adopt the following definition: point P belongs to the radiative zone if a photon 
(in the $\parallel$ polarization state) emitted parallel to $\bB$ with energy 
$\gsc\,b\,m_ec^2$ does not convert to a pair and escapes the 
magnetosphere. 

We have numerically calculated the boundary of the radiative zone for 
dipole magnetospheres around neutron stars with various magnetic 
moments and temperatures. The boundary is a closed surface around 
the star, which may be described 
in spherical coordinates by a function $r_0(\theta)$.
Instead of $r$ and $\theta$ we find it more convenient to use the 
coordinates $\Rmax$ and $b$, where $\Rmax$ labels the field lines and 
the value of $b$ is used as a coordinate running along the field line. Then 
the boundary is described by a function $b_0(\Rmax)$.  
The radiative zone only exists on sufficiently extended field lines,
\beq
    \Rmax>R_0\approx \left(\frac{\mum}{\BQ}\right)^{1/3}
                \left(\frac{\Q}{2}\right)^{1/6}
    \approx 
    5.4\times 10^6 \,\left(\frac{\mum}{10^{33} {\rm ~G~cm}^3}\right)^{1/3}
          \left(\frac{kT}{0.5\rm ~keV}\right)^{-1/6} {\rm cm}.
\eeq
The numerically calculated $b_0$ for $\Rmax>R_0$ is well approximated by
\beq
\label{eq:b0}
       b_0(\Rmax)=0.265\,\left(\frac{\mum}{10^{33} {\rm ~G~cm}^3}\right)^{0.15}
          \left(\frac{kT}{0.5\rm ~keV}\right)^{0.39} \,
                  \left(\frac{\Rmax}{\Rs}-1\right)^{1/4},
\eeq
where 
\beq
   \Rs=5\times 10^6 \,\left(\frac{\mum}{10^{33} {\rm ~G~cm}^3}\right)^{0.31}
          \left(\frac{kT}{0.5\rm ~keV}\right)^{-0.14} {\rm cm}.
\eeq
The corresponding location of the boundary in spherical coordinates, 
$r_0(\theta)$, is determined by the relations $\Rmax=r/\sin^2\theta$ and 
$B=(\mum/r^3)(1+3\cos^2\theta)^{1/2}$.
The exact boundary location is reproduced by \Eq~(\ref{eq:b0}) 
with the accuracy of a few percent
in a broad range of relevant $T$ ($0.2<kT<1$~keV) and 
$\mum>3\times 10^{31}$~G~cm$^3$.
Note that $\Rs$ appearing in \Eq~(\ref{eq:b0}) is slightly smaller than 
$R_0$. For the models shown in this paper, 
$\mum=5\times 10^{32}$~G~cm$^3$, $kT=0.5$~keV, and 
$\Rs\approx0.9R_0$.

We have also calculated the free path of scattered photons 
to absorption, $\labs$, in the pair-loading zone $b>b_0$. 
It depends on the scattering location 
and the scattering angle. Practically all 
photons scattered in the zone $b>b_0$ have $\labs/r\ll 1$, i.e. they
immediately convert to $e^\pm$ pairs after scattering. 
At the boundary $b=b_0$, there is a steep 
transition to the escape regime $\labs=\infty$. 
The small $\labs$ in the pair-loading zone implies that photons scattered on 
a field line $\Rmax$ convert to pairs on almost the same field line.
For the sample models presented in this paper, we found that the typical 
shift in $\Rmax$ before conversion is $\delta \Rmax/\Rmax\sim 0.1$ for 
photons scattered at $b=b_0$; $\delta \Rmax$ quickly decreases to 
essentially zero for photons scattered at $b>b_0$.
This fact simplifies calculations in \Sect~3, as one can assume that 
the processing of the outflow energy into $e^\pm$ pairs occurs 
independently on different field lines.

Note that $b_0$ significantly increases near the magnetic axis, where 
$u=R/\Rmax\rightarrow 0$; sufficiently close to the axis, the pair loading zone 
disappears. This fact is, however, of little importance, as the maximum
electric current, which corresponds to twist amplitude $\psi\sim 1$, 
is small near the axis ($I\propto u^2$, see \Eq~[\ref{eq:I}]), and its contribution 
to the observed luminosity is negligible. The luminosity peaks as $u^2$ 
at larger $u$; in this region $b_0\sim 1/4$.


\section{Emissivity of the j-bundle}

\Sect~3 described how the net emission from the j-bundle is obtained by
integrating along thin magnetic tubes and then taking the sum over the tubes. 
It is also useful to have an explicit expression for the spatial distribution of 
emission (e.g. in spherical coordinates) for a concrete magnetospheric
configuration. Below we write down the general expression for the emissivity 
per unit volume and then specialize to 
approximately dipole (moderately twisted) magnetospheres.

First note that a short piece $dl$ of a thin tube with magnetic flux $d\ff$ has 
volume
\beq
\label{eq:dV}
   dV=\frac{d\ff}{B}\,dl,
\eeq
where we used the fact that the perpendicular cross section of the flux tube 
equals $B^{-1}d\ff$. Then, using $dL/d\ff=\chi\Phi\,dI/d\ff$ (\Eq~\ref{eq:Lt})
and $dL/d\ff db=b_0^{-1}dL/d\ff$ (\Eq~\ref{eq:dLdb}), 
we find the produced luminosity per unit volume
\beq
\label{eq:dLdV}
   \frac{dL}{dV}=\frac{\chi\Phi}{2b_0}\,\left|\frac{dI}{d\ff}\,\frac{dB}{dl}\right|.
\eeq
Here the factor of 1/2 takes into account that $db$ corresponds to two 
pieces $dl$ of the tube, as the tube has two ``legs'' (two footpoints);
$\Phi$ is the voltage between the footpoints, i.e. the net voltage along 
the tube.
\Eq~(\ref{eq:dLdV}) shows that $dL/dV$ depends on 
the distribution of electric current among the magnetic field 
lines in the j-bundle; this distribution is given by
\beq
\label{eq:dIdf}
     \frac{dI}{d\ff}=\frac{j}{B}\approx\frac{c\,\psi}{4\pi \Rmax},
\eeq
where $j$ is the current density. In \Eq~(\ref{eq:dIdf})  we used the Maxwell 
equation $\nabla\times \bB=(4\pi/c)\bj$ with neglected displacement current,
and substituted $\nabla\times\bB\approx \psi B/\Rmax$. This relation is 
derived below for the twisted dipole magnetosphere; it remains approximately 
valid for more general configurations.
Substitution of \Eq~(\ref{eq:dIdf})
to \Eq~(\ref{eq:dLdV}) gives
\beq
  \frac{dL}{dV}=\frac{\chi\Phi}{b_0}\left|\frac{dB}{dl}\right|\,
                         \frac{c\psi}{8\pi\Rmax},
\eeq
where $\Rmax$ is the apex radius of the magnetic field line passing through 
$dV$.

Consider a dipole magnetosphere with a moderate axisymmetric twist. 
Its magnetic field is well approximated as the sum of a dipole poloidal field and 
a toroidal field $B_\phi$ (see B09). The poloidal flux function is that of a normal 
dipole,
\beq
   \ff(r,\theta)=2\pi\mum\,\frac{\sin^2\theta}{r}=\frac{2\pi\mum}{\Rmax}.
\eeq 
It is also convenient to use the dimensionless flux function
\beq
    u=\frac{\ff\,R}{2\pi\mum}=\frac{R}{\Rmax},
\eeq 
where $R$ is the radius of the neutron star and $\Rmax$ is the apex
radius of the magnetic field line. Variable $u<1$, like $\ff$, 
can be used to label the magnetic flux surfaces or magnetic field lines in the
axisymmetric magnetosphere.
Suppose the region $u<u_\star$ is activated, i.e. carries electric currents,
and has approximately uniform twist amplitude $\psi\approx\const$.
The electric current
flowing between the magnetic axis and a flux surface $u$ is given by (B09),
\beq
\label{eq:I}
   I(u)\approx \frac{c\mum\,\psi\,u^2}{4R^2}.
\eeq
It gives $dI/du$ and $dI/d\ff$ that confirms \Eq~(\ref{eq:dIdf}). One can also 
derive an explicit expression for $dB/dl$ that appears in \Eq~(\ref{eq:dLdV}).
For the approximately dipole field, $B=\mum\,r^{-3}(1+3\cos^2\theta)^{1/2}$ 
and 
\beq
\label{eq:dbdl}
  \frac{dB}{dl}=-\frac{3\mum\,(3+5\cos^2\theta)\cos\theta}{r^4\,(1+4\cos^2\theta)}.
\eeq

\end{appendix}



\end{document}